\documentclass[paper]{JHEP3}

\usepackage{epsfig,multicol}

\def\a{\alpha}
\def\b{\beta}
\def\c{\gamma}
\def\d{\delta}
\def\e{\epsilon}

\def\l{\lambda}
\def\m{\mu}
\def\n{\nu}

\def\r{\rho}

\def\t{\tau}

\def\C{\Gamma}
\def\D{\Delta}
\def\L{\Lambda}

\def\pl{\partial}

\def\rta{\rightarrow}

\def\la{\langle}
\def\ra{\rangle}

\def\Dslash{\,{\raise.15ex\hbox{/}\mkern-12mu D}}

\def\be{\begin{equation}}
\def\ee{\end{equation}}

\title{Pseudoscalar Meson Decay Constants and Couplings, 
the Witten-Veneziano Formula beyond large $N_c$, 
and the Topological Susceptibility
\thanks{This research is supported in part by 
PPARC grants PP/G/O/2002/00470 and PP/D507407/1.  }}

\author{G.M. Shore\\
        Department of Physics\\
        University of Wales, Swansea\\
        Swansea SA2 8PP, U.K.\\
        E-mail: \email{g.m.shore@swansea.ac.uk}}

\abstract{The QCD formulae for the radiative decays $\eta,\eta'\rightarrow\c\c$,
and the corresponding Dashen--Gell-Mann--Oakes--Renner relations, differ from 
conventional PCAC results due to the gluonic $U(1)_A$ axial anomaly. This 
introduces a critical dependence on the gluon topological susceptibility. 
In this paper, we revisit our earlier theoretical analysis of radiative 
pseudoscalar decays and the DGMOR relations and extract explicit experimental 
values for the decay constants. This is our main result. The flavour singlet 
DGMOR relation is the generalisation of the Witten-Veneziano formula beyond 
large $N_c$, so we are able to give a quantitative assessment of the 
realisation of the $1/N_c$ expansion in the $U(1)_A$ sector of QCD. 
Applications to other aspects of $\eta'$ physics, including the relation with 
the first moment sum rule for the polarised photon structure function $g_1^\c$, 
are highlighted. The $U(1)_A$ Goldberger-Treiman relation is extended to 
accommodate $SU(3)$ flavour breaking and the implications of a more precise
measurement of the $\eta$ and $\eta'$-nucleon couplings are discussed. 
A comparison with the existing literature on pseudoscalar meson decay constants 
using large-$N_c$ chiral Lagrangians is also made.
}

\begin{document}

\section{Introduction}

The phenomenology of the pseudoscalar mesons opens a window on many interesting
aspects of the non-perturbative dynamics of QCD, including spontaneous chiral
symmetry breaking, the electromagnetic and gluonic axial anomalies, the OZI
rule, the gluon topological susceptibility, and so on. Nevertheless, until
comparatively recently, phenomenological analyses did not take fully into
account the role of the gluonic $U(1)_A$ anomaly in this sector, with the result
that most of the existing determinations of the pseudoscalar meson decay
constants and couplings are based on over-simplified theoretical formulae
which miss the most interesting anomaly-sensitive physics.

In a previous paper \cite{Shore:1999tw} (see also \cite{Shore:1991pn}) 
we have analysed in detail the radiative decays
of the neutral pseudoscalar mesons, $\pi^0,\eta,\eta' \rta \c\c$,
and provided a set of formulae describing these processes together
with modified Dashen--Gell-Mann--Oakes--Renner (DGMOR)
\cite{Gell-Mann:1968rz,Dashen:1969eg} relations which fully
include the effect of the anomaly and the gluon topological susceptibility 
in the flavour singlet sector as well as explicit flavour $SU(3)$ breaking.
In this paper, we confront these formulae with experimental data to extract
a set of results for the four pseudoscalar meson decay constants,
$f^{0\eta'}, f^{0\eta}, f^{8\eta'}$ and $f^{8\eta}$.

Our approach, which is based on a straightforward generalisation of 
conventional PCAC to include the anomaly in a renormalisation group consistent
way, is unique in that it is theoretically consistent yet does rely on using 
the $1/N_c$ expansion. The decay formulae and DGMOR relations we derive are 
valid for all $N_c$. At some point, however, the fact that the $\eta'$ is not 
a Nambu-Goldstone (NG) boson means that to make predictions from our formulae
we need to augment the standard dynamical approximations of PCAC or
chiral perturbation theory with a further dynamical input. Here,
we make a judicious use of $1/N_c$ ideas to justify the use of the
lattice calculation of the topological susceptibility in pure Yang-Mills
theory as an input into our flavour singlet DGMOR relation. 
The self-consistency of this approach allows us to test the validity 
of $1/N_c$ methods in the $U(1)_A$ sector against real experimental data.
In particular, the flavour singlet DGMOR relation is the finite-$N_c$ 
generalisation of the well-known Witten--Veneziano formula
\cite{Witten:1979vv,Veneziano:1979ec}, and we can use 
our results to study how well the large-$N_c$ limit is realised in real QCD.  
This is especially important in view of the extensive use of large $N_c$ in
modern duality-based approaches to non-perturbative gauge theories such as 
the AdS-CFT correspondence.

The relation of our analysis of the low-energy decays $\eta'(\eta) \rta \c\c$
to high-energy physics in the form of the first moment sum rule for the 
polarised photon structure function $g_1^\c$ 
\cite{Narison:1992fd,Shore:1992pm} is also discussed. 
The dependence of this sum rule on the photon momentum encodes many aspects of 
non-perturbative QCD physics and its full measurement is now within the reach 
of planned high-luminosity $e^+ e^-$ colliders \cite{Shore:2004cb}. 
Our methods can also be readily extended to a variety of other reactions
involving the pseudoscalar mesons, including $\eta'(\eta)\rta V\c$,
where $V$ is one of the flavour singlet vector mesons $\rho,\omega,\phi$,~
$\eta'(\eta)\rta \pi^+ \pi^- \c$,~ a variety of electro and photoproduction
reactions such as $\c p \rta p \eta'(\eta)$~and $\c p \rta p \phi$,~
and $\eta$ and $\eta'$ production in $pp$ collisions, $pp \rta pp \eta'(\eta)$.

The latter are relevant for determining the couplings $g_{\eta NN}$ and
$g_{\eta' NN}$ which occur in generalisations of the Goldberger-Treiman
relation \cite{Goldberger:1958vp}. In this paper, we present a new version 
of the $U(1)_A$ Goldberger-Treiman relation 
\cite{Veneziano:1989ei,Shore:1990zu,Shore:1991dv} 
incorporating flavour octet-singlet mixing
in the $\eta - \eta'$ sector. We are then able to use our results for the 
decay constants to show how the physical interpretation of this relation, 
which is closely related to the `proton spin' problem in high-energy QCD
(i.e.~the first moment sum rule for the polarised proton structure
function $g_1^p$), depends critically on the experimental determination 
of $g_{\eta' NN}$.  

Finally, in an appendix, we give a brief comparison of the similarities
and differences between our approach and an alternative theoretically 
consistent framework for describing $\eta$ and $\eta'$ physics, 
viz.~the large-$N_c$ chiral Lagrangian formalism of 
refs.\cite{Leutwyler:1997yr,Kaiser:1998ds,Kaiser:2000gs} (see also
\cite{Gasser:1983yg,Herrera-Siklody:1996pm}). A comprehensive review of 
the phenomenology of the pseudoscalar mesons, incorporating the chiral 
Lagrangian results, is given in ref.\cite{Feldmann:1999uf}

\section{Radiative decay and DGMOR formulae}

In this section, we present the radiative decay and DGMOR formulae and 
discuss how to extract phenomenological quantities from them, emphasising
their $1/N_c$ dependence. A brief review of their derivation in the language
of `$U(1)_A$ PCAC' is given in section 3, but we refer to the original
papers \cite{Shore:1999tw,Shore:1991pn,Shore:2001cs} 
for full details, including a more precise formulation using
functional methods. Throughout, we consider the case of three light flavours,
$n_f=3$.

The assumptions made in deriving these formulae are the standard ones of PCAC.
The formulae are based on the zero-momentum chiral Ward identities. 
It is then assumed that the decay `constants' ($f^{a\a}(k^2)$ in the notation
below) and couplings ($g_{\eta^\a \c\c}(k^2)$) are approximately constant
functions of momentum in the range from zero, where the Ward identities
are applied, to the appropriate physical particle mass. It is important
that this approximation is applied only to pole-free quantities which have 
only an implicit dependence on the quark masses.  This is equivalent to
assuming pole-dominance of the propagators for the relevant operators
by the (pseudo-)NG bosons, and naturally becomes exact in the chiral
limit. Notice that these dynamical approximations are equally necessary
in the chiral Lagrangian approach, where they are built in to the
basic structure of the model -- the effective fields are
chosen to be in one-to-one correspondence with the NG bosons and 
pole-dominance is implemented through the low-momentum expansion with
momentum-independent parameters corresponding to the decay constants.  

The relations for the $\pi^0$ decouple from those for the $\eta$ and $\eta'$
in the realistic approximation of $SU(2)$ flavour invariance of the
quark condensates, i.e. $\la \bar u u \ra = \la \bar d d \ra$.  The decay 
formula for $\pi^0 \rta\c\c$ is the standard one,
\be
f_\pi g_{\pi\c\c} ~=~ a_{\rm em}^3 {\a_{\rm em}\over\pi}
\label{eq:ba}
\ee
with $a_{\rm em}^3 = {1\over3}N_c$, while the DGMOR formula is simply
\be
f_\pi^2 m_\pi^2 ~=~ - (m_u \la \bar u u \ra  + m_d \la \bar d d \ra)
\label{eq:bb}
\ee
The coupling $g_{\pi\c\c}$ is defined as usual from the decay amplitude:
\be
\la \c\c|\pi^0\ra ~=~ - i ~g_{\pi\c\c}~\e_{\l\r\a\b}p_1^\a p_2^\b 
\e^\l(p_1) \e^\r(p_2)
\label{eq:bc}
\ee
in obvious notation \cite{Shore:1999tw}. The r.h.s.~of eq.(\ref{eq:ba}) 
arises from the electromagnetic $U(1)_A$ anomaly and is of course sensitive
to the number of colours, $N_c$. The decay constant has a simple
theoretical interpretation as the coupling of the pion to the
flavour triplet axial current, $\la 0|J_{\m 5}^3|\pi\ra = i k_\m f_\pi$,
which is conserved in the chiral limit. 

In the $\eta,\eta'$ sector, however, explicit $SU(3)$ flavour breaking
means there is mixing. The decay constants therefore form a $2\times 2$
matrix:
\be
f^{a\a} ~=~ \left(\matrix{f^{0\eta'} &f^{0\eta}\cr f^{8\eta'} &f^{8\eta}}
\right)
\label{eq:bd}
\ee
where the index $a$ is an $SU(3)\times U(1)$ flavour index 
(including the singlet $a=0$)
and $\a$ denotes the physical particle states $\pi^0,\eta,\eta'$. 
These four decay constants are independent 
\cite{Leutwyler:1997yr,Kaiser:1998ds,Feldmann:1997vc,Feldmann:1998vh}, 
in contradiction to the original phenomenological parametrisations 
which erroneously expressed $f^{a\a}$ as a diagonal decay constant 
matrix times an orthogonal $\eta -\eta'$ mixing matrix, giving a total 
of only three independent parameters.  Sometimes the four decay constants 
are expressed in terms of two constants and two mixing angles but, 
while we also express our results in this form in section 4, there is 
no particular reason to parametrise in this way.

The other new feature in the $\eta,\eta'$ sector is the presence of the
gluonic $U(1)_A$ anomaly in the flavour singlet current. This means that
even in the chiral limit the $\eta'$ is not a true NG boson and 
therefore the direct analogue of the PCAC formula (\ref{eq:ba}) does not 
hold. Nonetheless, we can still write an analogous formula but with an 
extra term related to the gluon topological susceptibility. 

The decay formulae are\footnote{Notice that compared to 
refs.\cite{Shore:1999tw,Shore:1991pn,Shore:2001cs,Shore:2004cb}
we have changed the normalisation of the flavour singlets by a factor
$1/\sqrt{2n_f}= 1/\sqrt6$. The normalisations here are ${\rm tr} T^a T^b = 
{1\over2}\d_{ab}$ for generators $T^a$ ($a = 0,1,\ldots 8$) of 
$SU(3)\times U(1)$, i.e.~$T^i = {1\over2}\l^i$ ($i = 1,\ldots 8$) where 
$\l^i$ are the Gell-Mann matrices, and $T^0 = {1\over\sqrt6}{\bf 1}$ for
the flavour singlet. The $d$-symbols are defined by $\{T^a,T^b\} =
d_{abc}T^c$ and include $d_{000} = d_{033} = d_{088} =
d_{330} = d_{880} = \sqrt{2\over3}$, $d_{338} = d_{383} = -d_{888} =
\sqrt{1\over3}$.}:
\begin{eqnarray}
f^{0\eta'} g_{\eta'\c\c} + f^{0\eta} g_{\eta\c\c} + {\sqrt6} A g_{G\c\c} 
~~=~~a_{\rm em}^0 {\a_{\rm em}\over\pi}
\label{eq:be}\\
\nonumber\\
f^{8\eta'}g_{\eta'\c\c} + f^{8\eta} g_{\eta\c\c} 
~~=~~a_{\rm em}^8 {\a_{\rm em}\over\pi}
\label{eq:bf}\\ 
\nonumber
\end{eqnarray}
where $a_{\rm em}^0 = {2\sqrt2\over3\sqrt3}N_c$ and 
$a_{\rm em}^8 = {1\over3\sqrt3}N_c$,
and the corresponding DGMOR relations are:
\begin{eqnarray}
&(f^{0\eta'})^2 m_{\eta'}^2 + (f^{0\eta})^2 m_\eta^2 ~~=~~ 
- {2\over3}\bigl(m_u \la\bar u u\ra + m_d \la\bar d d\ra 
+ m_s \la\bar s s\ra \bigr)  + 6 A 
\label{eq:bg}\\
\nonumber\\
&f^{0\eta'} f^{8\eta'} m_\eta'^2 + f^{0\eta} f^{8\eta} m_{\eta}^2 ~~=~~ 
- {\sqrt2\over3} \bigl(m_u \la\bar u u\ra + m_d \la\bar d d\ra 
- 2 m_s \la\bar s s\ra \bigr)
\label{eq:bh}\\
\nonumber\\
&(f^{8\eta'})^2 m_\eta'^2 + (f^{8\eta})^2 m_{\eta}^2 ~~=~~ 
-{1\over3}\bigl(m_u \la\bar u u\ra + m_d \la\bar d d\ra +  
4 m_s \la\bar s s\ra \bigr)
\label{eq:bi}\\
\nonumber
\end{eqnarray}

There are several features of these equations which need to be explained.
First, because of the anomaly, the decay constants in the flavour
singlet sector can {\it not} be identified as the couplings of the 
$\eta,\eta'$ to the singlet axial-vector current. In particular, the object 
$F^0_{\eta'}$ defined as $\la 0|J_{\m 5}^0|\eta'\ra = i k_\m F^0_{\eta'}$
is not a renormalisation group invariant and so is not to be identified
as a physical decay constant. The derivation of the radiative decay and
DGMOR formulae in the next section makes it clear that $F^0_{\eta'}$
plays no role -- certainly it is {\it not} $f^{0\eta'}$.  
Instead, our decay constants are defined in terms of the couplings of the 
$\pi,\eta,\eta'$ to the pseudoscalar currents through the relations 
$f^{a\a}\la 0| \phi_5^b |\eta^\a\ra = d_{abc}\la\phi^c\ra$ (for notation, 
see section 3). This coincides with the usual definition except in the
flavour singlet case.

The coefficient $A$ appearing in the flavour singlet equations is the
non-perturbative constant that determines the topological susceptibility
in QCD. Recall that the topological susceptibility $\chi(0)$ is 
defined as
\be
\chi(0) ~=~ \int d^4 x ~i \la 0|T~Q(x)~Q(0)|0\ra
\label{eq:bj}
\ee
where $Q = {\a_s\over8\pi}{\rm tr}G^{\m\n}\tilde G_{\m\n}$ is the gluon
topological charge. The anomalous chiral Ward identities determine the
dependence of $\chi(0)$ on the quark masses and condensates up to a 
single non-perturbative parameter \cite{DiVecchia:1980ve,Shore:1999tw}, viz:
\be
\chi(0) ~~=~~-A \biggl(1 - A \sum_{q=u,d,s}{1\over m_q\la\bar q q\ra}
\biggr)^{-1} 
\label{eq:bk}
\ee
or alternatively,
\begin{eqnarray}
\chi(0) ~~=~~ {- A~ m_u m_d m_s \la\bar u u\ra \la\bar d d\ra \la\bar s s\ra 
\over m_u m_d m_s \la\bar u u\ra \la\bar d d \ra \la\bar s s \ra
-A\bigl(m_u m_d \la\bar u u\ra \la\bar d d \ra + m_u m_s \la\bar u u\ra 
\la\bar s s \ra + m_d m_s \la\bar d d \ra \la\bar s s \ra \bigr)}\nonumber\\
\label{eq:bl}
\end{eqnarray}
Notice how the well-known result that $\chi(0) = 0$ if any of the quark
masses is zero is realised in this expression.

The final new element in the flavour singlet decay formula is the coupling 
parameter $g_{G\c\c}$.  This is unique to our approach. It takes
account of the fact that, because of mixing with the pseudoscalar gluon
operator $Q$ due to the anomaly, the physical $\eta'$ is not a NG boson.
$g_{G\c\c}$ is {\it not} a physical coupling, although it may reasonably be 
thought of as the coupling of the photons to the gluonic component of the 
$\eta'$. (A justification for this picture is given later.) 
Its precise field-theoretic definition is given in section 3 and 
refs.\cite{Shore:1999tw,Shore:1991pn,Shore:2001cs}. 
We should emphasise, however, that an analogous parameter
must appear in {\it any} correct current algebra analysis involving the
$\eta'$, such as the $1/N_c$ chiral Lagrangian approach of 
refs.\cite{Leutwyler:1997yr,Kaiser:1998ds,Kaiser:2000gs}.
Phenomenologically, it may be found to be relatively small (we test this
numerically in section 4), but any formulae which leave out such a term
completely are inevitably theoretically inconsistent.

Of course, in addition to these formulae, we also have the DGMOR relations
for the remaining pseudoscalars. For example, for the $K^+$,
\be
f_K^2 m_K^2 ~=~ - (m_u \la \bar u u \ra  + m_s \la \bar s s \ra)
\label{eq:bm}
\ee
If we assume exact $SU(2)$ flavour symmetry, we can then use 
eqs.(\ref{eq:bb}) and (\ref{eq:bm}) to eliminate the quark masses and
condensates in favour of $f_\pi, f_K, m_\pi^2$ and $m_K^2$ in the DGMOR
formulae for the $\eta$ and $\eta'$. We find:
\begin{eqnarray}
&(f^{0\eta'})^2 m_{\eta'}^2 + (f^{0\eta})^2 m_\eta^2 ~~=~~ 
{1\over3} \bigl(f_\pi^2 m_\pi^2 + 2 f_K^2 m_K^2\bigr) + 6 A 
\label{eq:bn}\\
\nonumber\\
&f^{0\eta'} f^{8\eta'} m_\eta'^2 + f^{0\eta} f^{8\eta} m_{\eta}^2 ~~=~~ 
{2\sqrt2\over3}\bigl(f_\pi^2 m_\pi^2 - f_K^2 m_K^2\bigr)
\label{eq:bo}\\
\nonumber\\
&(f^{8\eta'})^2 m_\eta'^2 + (f^{8\eta})^2 m_{\eta}^2 ~~=~~ 
-{1\over3}\bigl(f_\pi^2 m_\pi^2 - 4 f_K^2 m_K^2\bigr)
\label{eq:bp}\\
\nonumber
\end{eqnarray}

A quick look at the full set of decay and DGMOR formulae 
(\ref{eq:be}),(\ref{eq:bf}) and (\ref{eq:bn})--(\ref{eq:bp}) 
now shows that we have five equations for six parameters,
viz. $f^{0\eta'}, f^{0\eta}, f^{8\eta'}, f^{8\eta}, A$ and $g_{G\c\c}$,
assuming that $f_\pi$, $f_K$ and the physical masses are known along with the
experimental values for the couplings $g_{\eta'\c\c}$ and $g_{\eta\c\c}$.
It is not surprising that this set is under-determined. In particular,
the necessary presence of the unphysical coupling $g_{G\c\c}$ in the flavour 
singlet decay equation essentially removes its predictivity. 
At this stage, the best we can do is to evaluate the singlet decay
constants and the gluonic coupling $g_{G\c\c}$ as functions of the
topological susceptibility parameter $A$. This is done in section 5.
In order to make more progress, we therefore need a further, dynamical input.

As yet, everything we have done has been entirely independent of the
$1/N_c$ expansion. The $1/N_c$ expansion (or the OZI limit -- see 
ref.\cite{Veneziano:1990mx} 
for a careful discussion of the differences) is known to 
give a good approximation to many aspects of the dynamics of QCD and
could provide the required extra input, but its application to the 
$U(1)_A$ sector needs to be handled with great care. For example, in the
chiral limit, the mass of the $\eta'$, which arises due to the anomaly,
is formally $m_{\eta'}^2 = O(1/N_c)$; however, numerically (allowing
for the quark masses going to zero) this is of the same order of magnitude
as a typical meson such as the $\r$ and is certainly not small. 
Generally, it is not clear that quantities which are formally suppressed 
in $1/N_c$ are in fact numerically suppressed in real QCD, so we must be 
extremely careful in applying $1/N_c$ methods here. 

Despite these caveats, we will see that $1/N_c$ can play a useful role
in analysing the decay formulae and DGMOR relations.  Conventional large 
$N_c$ counting gives the following orders for the various quantities:
$f^{a\a} = O(\sqrt N_c)$ for all the decay constants; 
the $\eta$ and $\eta'$ couplings $g_{\eta^\a \c\c} = O(\sqrt N_c)$
but $g_{G\c\c} = O(1)$; $m_{\eta'}^2$ and $m_{\eta}^2$ are both $O(1)$, but
note that $m_{\eta'}^2 = O(1/N_c)$ in the chiral limit -- the numerically
dominant contribution to its mass from the anomaly is formally $1/N_c$
suppressed relative to the $O(1)$ contribution from the explicit
chiral symmetry breaking quark masses; the condensates $\la \bar q q\ra
= O(N_c)$; the anomaly coefficients $a_{\rm em}^a = O(N_c)$;
while finally the coefficient in the topological susceptibility is
$A = O(1)$.

Referring to eqs.(\ref{eq:bk}) or (\ref{eq:bl}), we therefore see that at 
large $N_c$, $\chi(0) \simeq -A = O(1)$. Moreover, it is clear from
looking at planar diagrams that at leading order in $1/N_c$, $\chi(0)$
in QCD coincides with the topological susceptibility in pure Yang-Mills
theory, $\chi(0)|_{YM}$.  It follows that 
\be
A ~=~ \chi(0)|_{YM} + O(1/N_c)
\label{eq:bq}
\ee
a result that plays an important role in the Witten-Veneziano formula
(see below).

Now consider the flavour singlet DGMOR relation (\ref{eq:bg}) or 
(\ref{eq:bn}). Each term is $O(N_c)$ apart from $A$, which is $O(1)$.
Naively, we might think that this sub-leading term would be small
so could be neglected. However, we know that it contributes at the same
order as the sub-leading term in $(f^{0\eta'})^2 m_{\eta'}^2$ given
by the large anomaly-induced $O(1/N_c)$ contribution to the $\eta'$
mass squared. So the topological susceptibility contribution to
this relation is crucial, even though it is formally suppressed in $1/N_c$.
However, we may reasonably expect that the further $O(1/N_c)$ 
corrections to $A$ are genuinely small and that a good numerical
approximation to eqs.(\ref{eq:bg}) or (\ref{eq:bn}) is obtained by
keeping only the terms up to $O(1)$. This means that we may sensibly
approximate the parameter $A$ by $\chi(0)|_{YM}$ in the DGMOR relation.

This is the crucial simplification. A reliable estimate of $\chi(0)|_{YM}$
at $O(1)$ is available from lattice calculations in pure Yang-Mills
theory \cite{DelDebbio:2004ns}. This extra dynamical input allows us to 
determine the four decay constants from eqs.(\ref{eq:bf}) and 
(\ref{eq:bn})--(\ref{eq:bp}). We can then analyse the flavour singlet 
decay formula (\ref{eq:be}) to determine $g_{G\c\c}$ and see how important 
this new term actually is numerically.

We already know that the $A g_{G\c\c}$ contribution to eq.(\ref{eq:bl})
is suppressed by one power of $1/N_c$ compared to the other terms
in the formula. Moreover, $g_{G\c\c}$ is renormalisation group invariant
(see refs.\cite{Shore:1991pn,Shore:2001cs}). 
Now, in previous work we have developed 
an intuition as to when it is likely to be reliable to assume that
$1/N_c$  suppressed terms are actually numerically small. The
argument is based on the idea that violations of the OZI rule
are associated with the $U(1)_A$ anomaly, so that we can identify 
OZI-violating quantities by their dependence on the anomalous dimension
associated with the non-trivial renormalisation of $J_{\m 5}^0$ due
to the anomaly. RG non-invariance can therefore be used as an indicator
of which quantities we expect to show large OZI violations.
In this case, $g_{G\c\c}$ is RG invariant, so we would expect the
OZI rule to be good. This means that since it is OZI suppressed
(essentially, higher order in $1/N_c$) relative to the $\eta'$
decay coupling $g_{\eta^\a \c\c}$, it should be numerically smaller
as well.\footnote{To be precise, the conjecture is that the contribution
$\sqrt6 A g_{G\c\c}$ in the decay formula will be small compared
to the dominant term $f^{0\eta'} g_{\eta'\c\c}$.  Note that as defined
the dimensions of $g_{G\c\c}$ and $g_{\eta'\c\c}$ are different, so they 
can not be directly compared.}

We will test this conjecture with the experimental data in section 4.
The issue is an important one.  If $g_{G\c\c}$ can be neglected, then the 
naive current algebra formulae will turn out phenomenologically to be a 
good approximation to data, even though they are theoretically inconsistent.
This would apply not just to the radiative pseudoscalar decays, but to
a whole range of current algebra processes involving the $\eta'$
(see, for example, refs.\cite{Feldmann:1999uf,Shore:2001cs}). 
We have also used this conjecture in our analysis of the closely 
related `proton spin' problem in polarised deep inelastic scattering where 
it plays an important role in our prediction of the first moment of the 
polarised proton structure function $g_1^p$ \cite{Narison:1998aq}. 

Finally, to close this section, we show in detail how these DGMOR
relations are related to the well-known Witten-Veneziano formula
for the mass of the $\eta'$, which is derived in the large-$N_c$ limit
of QCD. In fact, this is simply the $N_c \rta \infty$ limit
of the flavour singlet DGMOR formula, which we emphasise is valid 
for all $N_c$. To see this in detail, recall that the Witten-Veneziano 
formula for non-vanishing quark masses is \cite{Veneziano:1979ec}
\be
m_{\eta'}^2 + m_{\eta}^2 - 2 m_K^2 ~=~ - {6\over f_\pi^2} \chi(0)|_{YM}
\label{eq:br}
\ee
Of course, only the $m_{\eta'}^2$ term on the l.h.s.~is present in the 
chiral limit.
Now add the DGMOR formulae (\ref{eq:bn}) and (\ref{eq:bp}). We find
\be
(f^{0\eta'})^2 m_{\eta'}^2 + (f^{0\eta})^2 m_\eta^2 +
(f^{8\eta})^2 m_\eta^2 + (f^{8\eta'})^2 m_{\eta'}^2 - 2 f_K^2 m_K^2 
~~=~~ 6A
\label{eq:bs}
\ee
To reduce this to its Witten-Veneziano approximation, we impose the 
large-$N_c$ limit to identify the full QCD topological charge parameter 
$A$ with $-\chi(0)|_{YM}$ according to eq.(\ref{eq:bq}). We then set the
`mixed' decay constants $f^{0\eta}$ and $f^{8\eta'}$ to zero
and all the remaining decay constants $f^{0\eta'}, f^{8\eta}$ and
$f_K$ equal to $f_\pi$. With these approximations, we recover 
eq.(\ref{eq:br}).  

In section 4, when we find the explicit experimental values for all
these quantities in real QCD, we will be able to judge how good an
approximation the large-$N_c$ Witten-Veneziano formula is to our
general DGMOR relation.

\section{Theory}

We now sketch the `$U(1)_A$ PCAC' derivation of the decay formulae and
DGMOR relations. For a more precise treatment in terms of functional
chiral Ward identities and a complete renormalisation group analysis,
as well as an effective Lagrangian formulation, we refer to our
original papers \cite{Shore:1999tw,Shore:1991pn,Shore:2001cs}.

The starting point is the $U(1)_A$ chiral anomaly equation in pure QCD
with $n_f$ flavours of massive quarks, viz\footnote{We use the following
$SU(3)$ notation for the quark masses and condensates:
$$
\left(\matrix{m_u&0&0\cr 0&m_d&0\cr 0&0&m_s\cr}\right)
~=~ \sum_{a=0,3,8} m^a T^a
$$
and
$$
\left(\matrix{\la\bar u u\ra &0 &0 \cr 0 &\la\bar d d\ra &0\cr
0 &0 &\la\bar s s\ra\cr}\right) 
~=~ 2 \sum_{a=0,3,8} \la\phi^a\ra T^a
$$
where $\la\phi^a\ra$ is the VEV of $\phi^a = \bar q T^a q$. It is also very
convenient to introduce the compressed notation
$$
M_{ab} = d_{acb} m^c ~~~~~~~~~~~~~~~~~~~ \Phi_{ab} = d_{abc}\la\phi^c\ra
$$}
\be
\pl^\m J_{\m 5}^a ~=~ M_{ab} \phi_5^b + \sqrt{2n_f} Q \d_{a0}
\label{eq:ca}
\ee
where the axial vector current is $J_{\m 5}^a = \bar q \c_\m \c_5 T^a q$
and the pseudoscalar is $\phi_5^a = \bar q \c_5 T^a q$.
The corresponding chiral Ward identities for the two-point Green 
functions of interest are therefore (in momentum space):
\begin{eqnarray}
ik^\m \la J_{\m 5}^a~Q \ra - \sqrt{2n_f} \d_{a0} \la Q~Q \ra
- M_{ac} \la \phi_5^c~Q\ra ~~=~~0 
\label{eq:cb}\\
ik^\m \la J_{\m 5}^a~\phi_5^b \ra - \sqrt{2n_f} \d_{a0} \la Q~\phi_5^b \ra
- M_{ac} \la \phi_5^c~\phi_5^b\ra ~~=~~ \Phi_{ab}
\label{eq:cc}\\
\nonumber
\end{eqnarray}

Since there are no massless pseudoscalar mesons in the theory, the
terms in these identities involving the axial current vanish at 
zero momentum because of the explicit factors of $k^\m$. The zero-momentum
chiral Ward identities simply comprise the remaining terms. In
particular, we have
\be
M_{ac}M_{bd} \la \phi_5^b~\phi_5^d \ra = - (M\Phi)_{ab}
+ (2n_f) \la Q~Q \ra \d_{a0} \d_{b0}
\label{eq:cd}
\ee
We can also derive the result quoted in eq.(\ref{eq:bk}) for the 
topological susceptibility. In this notation,
\be
\chi(0) ~\equiv~ \la Q~Q \ra ~=~
{-A\over 1 - (2n_f) A (M\Phi)_{00}^{-1}}
\label{eq:ce}
\ee

The physical mesons $\eta^\a = (\pi^0,\eta^0,\eta')$ couple to the
pseudoscalar operators $\phi_5^a$ and $Q$, so their properties
may be deduced from the two-point Green functions above. In the PCAC
approximation, as explained in section 2, the zero-momentum Ward
identities are sufficient. In order to make the correspondence between
the QCD operators and the physical mesons as close as possible, it
is convenient to redefine linear combinations such that the propagator 
(two-point Green function) matrix is diagonal and properly normalised.
We therefore define operators $\eta^a$ and $G$ such that 
\be
\left(\matrix{\la Q~Q \ra & \la Q~\phi_5^b \ra \cr \la \phi_5^a ~ Q 
&\la \phi_5^a~\phi_5^b \ra}\right) ~~\rta~~
\left(\matrix{\la G~G \ra &0 \cr 0 &\la \eta^\a ~\eta^\b \ra \cr}\right)
\label{eq:cf}
\ee
This is achieved by taking
\be
G ~=~ Q - \la Q~\phi_5^a \ra \bigl(\la \phi_5~\phi_5\ra\bigr)_{ab}^{-1}
\phi_5^b
\label{eq:cg}
\ee
which reduces at zero momentum to 
\be
G ~=~ Q + 2n_f A \Phi_{0b}^{-1} \phi_5^b
\label{eq:ch}
\ee
and defining
\be
\eta^\a ~=~ f^{T\a a} \Phi_{ab}^{-1} \phi_5^b
\label{eq:ci}
\ee
With this choice, the $\la G~G \ra$ propagator at zero momentum is
simply
\be
\la G~G \ra ~=~ -A
\label{eq:cj}
\ee
and we demand that the $\la\eta^\a~\eta^\b\ra$ propagator has
the canonical normalisation
\be
\la \eta^\a~\eta^\b \ra ~=~ {1\over k^2 - m_{\eta^\a}^2} \d^{\a\b}
\label{eq:ck}
\ee
This normalisation fixes the decay constants introduced in eq.(\ref{eq:ci}).
The DGMOR relations then follow immediately from the zero-momentum 
chiral Ward identity (\ref{eq:cd}) and the expression (\ref{eq:ce}) for the
topological susceptibility. We find
\begin{eqnarray}
f^{a\a} m_{\a\b}^2 f^{T\b b} ~~=~~ 
&\Phi_{ac} \bigl(\la \phi_5~\phi_5 \ra\bigr)_{cd}^{-1} \Phi_{db} 
\nonumber\\
{}\nonumber\\
=~~ &- (M\Phi)_{ab} + (2n_f) A \d_{a0} \d_{b0}
\label{eq:cl}
\end{eqnarray}
Unwrapping the condensed $SU(3)$ notation then shows that this matrix
equation is simply the set of DGMOR relations (\ref{eq:bg})--(\ref{eq:bi}).

The next step is the PCAC calculation of $\eta^\a \rta \c\c$. We implement
PCAC by the identification
\be
\pl^\m J_{\m 5}^a ~\rta~ f^{a\a} m_{\a\b}^2 \eta^\b + \sqrt{2n_f} G \d_{a0}
\label{eq:cm}
\ee
which follows from the anomaly eq.(\ref{eq:ca}) and the definitions of
the fields $G$ and $\eta^\a$. To be precise, the $\rta$ notation in 
eq.(\ref{eq:cm}) means that the identification can be made for insertions 
of the operators into zero-momentum Green functions and matrix elements only.
Notice that it is {\it not} valid at non-zero momentum, in particular for
on-shell matrix elements. It is the natural generalisation of the 
familiar PCAC relation $\pl^\m J_{\m 5}^3 \rta f_\pi m_\pi^2 \pi$, defining
the phenomenological pion field $\pi$. 

The other input is the full axial anomaly for QCD coupled to electromagnetism,
viz.
\be
\pl^\m J_{\m 5}^a ~=~ M_{ab} \phi_5^b + \sqrt{2n_f} Q \d_{a0} + a_{\rm em}^a
{\a\over8\pi} F^{\m\n} \tilde F_{\m\n}
\label{eq:cn}
\ee
where $a_{\rm em}^a$ are the anomaly coefficients given in section 2.

Implementing the PCAC relation (\ref{eq:cm}) together with the full anomaly
equation, we therefore find
\begin{eqnarray}
ik^\m \la\c\c|J_{\m 5}^a|0\ra ~~=~~ f^{a\a} m_{\a\b}^2 \la\c\c|\eta^\b|0\ra
+ \sqrt{2n_f} \la\c\c|G|0\ra \d_{a0} + a_{\rm em}^a {\a\over8\pi} 
\la\c\c|F^{\m\n} \tilde F_{\m\n}|0\ra \nonumber\\
{}\nonumber\\
= f^{a\a} m_{\a\b}^2 \la\eta^\b~\eta^\c\ra \la\c\c|\eta^\c\ra
+ \sqrt{2n_f} \la G~G\ra \la\c\c|G\ra \d_{a0} + a_{\rm em}^a {\a\over8\pi} 
\la\c\c|F^{\m\n} \tilde F_{\m\n}|0\ra \nonumber\\  
\label{eq:co}
\end{eqnarray}
at zero momentum, where we have used the fact that the propagators are
diagonal in the $\eta^\a, G$ basis. The l.h.s. vanishes at zero momentum
as there is no massless particle coupling to the axial current. Then,
using the explicit expressions (\ref{eq:cj}) and (\ref{eq:ck}) for the $G$ 
and $\eta^\a$ propagators, we find the decay constant formulae:
\be
f^{a\a} g_{\eta^\a \c\c} + \sqrt{2n_f} A g_{G\c\c} \d_{a0} ~~=~~ a_{\rm em}^a 
{\a\over\pi}
\label{eq:cp}
\ee
The novel coupling $g_{G\c\c}$ is precisely defined from the matrix element
$\la\c\c|G\ra$ in analogy with eq.(\ref{eq:bc}) for the conventional couplings.

Finally, notice that the mixing of states is conjugate to the mixing of fields.
In particular, the mixing for the states corresponding to eqs.(\ref{eq:ch})
and (\ref{eq:ci}) for the fields $G$ and $\eta^\a$ is
\be
|G\ra ~=~ |Q\ra
\label{eq:cq}
\ee
and
\be
|\eta^\a\ra ~=~  (f^{-1})^{\a a} \bigl(\Phi_{ab} |\phi_5^b\ra -
\sqrt{2n_f} A \d_{a0} |Q\ra \bigr)
\label{eq:cr}
\ee 
In this sense, we see that we can regard the physical $\eta'$ (and, with
$SU(3)$ breaking, the $\eta$) as an admixture of quark and gluon 
components, while the unphysical state $|G\ra$ is purely gluonic.
This is why we can usefully picture the unphysical coupling $g_{G\c\c}$
as the coupling of the photons to the anomaly-induced gluonic component
of the $\eta'$, as already mentioned in section 2.

\section{Phenomenology}

In this section, we use the experimental data on the radiative
decays $\eta,\eta' \rta \c\c$ to deduce values for the 
pseudoscalar meson decay constants $f^{0\eta'}$, $f^{0\eta}$, $f^{8\eta'}$ 
and $f^{8\eta}$ from the set of decay formulae (\ref{eq:be}),(\ref{eq:bf})
and DGMOR relations (\ref{eq:bn})-(\ref{eq:bp}). We will also find the 
value of the unphysical coupling parameter $g_{G\c\c}$ and test the
realisation of the $1/N_c$ expansion in real QCD.

The two-photon decay widths are given by
\be
\C\bigl(\eta'(\eta)\rta\c\c\bigr) ~~=~~ {m_{\eta'(\eta)}^3 \over64\pi~}
|g_{\eta'(\eta)\c\c}|^2
\label{eq:da}
\ee
The current experimental data, quoted in the Particle Data Group
tables \cite{PDG}, are
\be
\C(\eta'\rta\c\c) ~~=~~ 4.28 \pm 0.19 ~{\rm KeV}
\label{eq:db}
\ee
which is dominated by the 1998 L3 data \cite{L3} on the two-photon
formation of the $\eta'$ in $e^+ e^- \rta e^+ e^- \pi^+ \pi^- \c$,
and
\be
\C(\eta\rta\c\c) ~~=~~ 0.510 \pm 0.026 ~{\rm KeV}
\label{eq:dc}
\ee
which arises principally from the 1988 Crystal Ball \cite{Crystal} and
1990 ASP \cite{ASP} results on $e^+ e^- \rta e^+ e^- \eta$. (Notice that
we follow the note in the 1994 PDG compilation \cite{PDG1994} and use only 
the two-photon $\eta$ production data.)

From this data, we deduce the following results for the couplings 
$g_{\eta'\c\c}$ and $g_{\eta\c\c}$:
\be
g_{\eta'\c\c} ~~=~~ 0.031 \pm 0.001 ~{\rm GeV}^{-1}
\label{eq:dd}
\ee
and
\be
g_{\eta\c\c} ~~=~~ 0.025 \pm 0.001 ~{\rm GeV}^{-1}
\label{eq:de}
\ee
which may be compared with $g_{\pi\c\c} = 0.024 \pm 0.001 ~{\rm GeV}$.

We also require the pseudoscalar meson masses:
\begin{eqnarray}
m_{\eta'} ~=~ 957.78 \pm 0.14 ~{\rm MeV} ~~~~~~~~ 
m_{\eta}  ~=~ 547.30 \pm 0.12 ~{\rm MeV} \nonumber\\
m_K       ~=~ 493.68 \pm 0.02 ~{\rm Mev} ~~~~~~~~
m_\pi     ~=~ 139.57 \pm 0.00 ~{\rm MeV}
\label{eq:df}
\end{eqnarray}
and the decay constants $f_\pi$ and $f_K$. These are defined in the
standard way, so we take the following values (in our normalisations) 
from the PDG \cite{PDG}:
\be
f_K ~=~ 113.00 \pm 1.03 ~{\rm MeV} ~~~~~~~~
f_\pi ~=~ 92.42 \pm 0.26 ~{\rm MeV}
\label{eq:dg}
\ee
giving $f_K/f_\pi = 1.223 \pm 0.012$.

The final input, as explained in section 2, is the lattice calculation
of the topological susceptibility in pure Yang-Mills theory.
The most recent value, obtained in ref.\cite{DelDebbio:2004ns}, is
\be
\chi(0)|_{YM} ~~=~~ -(191 \pm 5 ~{\rm MeV})^4 ~~=~~ 
-(1.33 \pm 0.14) \times 10^{-3} ~{\rm GeV}^4
\label{eq:dh}
\ee
This supersedes the original value $\chi(0)|_{YM} \simeq -(180~{\rm MeV})^4$
obtained some time ago \cite{DiGiacomo:1990ij}. 
Similar estimates are also obtained 
using QCD spectral sum rule methods \cite{Narison:1990cz}. 
Using the argument explained in section 2, we therefore adopt the value
\be
A ~~=~~ (1.33 \pm 0.14) \times 10^{-3} ~{\rm GeV}^4
\label{eq:di}
\ee
for the non-perturbative parameter determining the topological susceptibility
in full QCD. 

The strategy is now to solve the set of five simultaneous equations
(\ref{eq:be}),(\ref{eq:bf}) and (\ref{eq:bn}),(\ref{eq:bo}), (\ref{eq:bp})
for the five remaining unknowns $f^{0\eta'}$, $f^{0\eta}$, $f^{8\eta'}$, 
$f^{\eta}$ and $g_{G\c\c}$. The results are\footnote{Note that this analysis 
includes the errors from the experimental inputs and the lattice evaluation
of $\chi(0)|_{YM}$, but {\it not} the systematic effect of the approximation
$A = -\chi(0)|_{YM}$.  The errors on the singlet decay constants are
dominated by the error on $A$. Isolating this, we have
$$
f^{0\eta'} = 104.2 \pm 0.3 \pm 4.0 ~{\rm MeV}~~~~~~~~
f^{0\eta} = 22.8 \pm 3.5 \pm 4.5 ~{\rm MeV}
$$
The octet decay constants are of course unaffected by the value of $A$.}:
\begin{eqnarray}
{}\nonumber\\
f^{0\eta'} ~=~ 104.2 \pm 4.0 ~{\rm MeV} ~~~~~~~~
f^{0\eta}  ~=~ 22.8 \pm 5.7 ~{\rm MeV} \nonumber\\
f^{8\eta'} ~=~-36.1 \pm 1.2 ~{\rm MeV} ~~~~~~~~
f^{8\eta}  ~=~ 98.4 \pm 1.4 ~{\rm MeV}
\label{eq:dj}
\end{eqnarray}
that is
\begin{eqnarray}
{f^{0\eta'}\over f_\pi} ~=~ 1.13 \pm 0.04   ~~~~~~~~
{f^{0\eta}\over f_\pi}  ~=~ 0.25 \pm 0.06 \nonumber\\
{f^{8\eta'}\over f_\pi} ~=~-0.39 \pm 0.01   ~~~~~~~~
{f^{8\eta}\over f_\pi}  ~=~ 1.07 \pm 0.02
\label{eq:dk}
\end{eqnarray}
and
\be
g_{G\c\c} ~=~ - 0.001 \pm 0.072 ~{\rm GeV}^{-4}
\label{eq:dl}
\ee
These are the main results of this paper. 

Before going on to consider their significance, we can re-express the
results for the decay constants in terms of one of the two-angle 
parametrisations used in the literature. We reiterate that in our view
there is no particular virtue in parametrising in this way,
but in order to help comparison with other analyses it may be helpful 
to present our results in this form as well. 
Refs.\cite{Leutwyler:1997yr,Kaiser:1998ds} define 
\be
\left(\matrix{f^{0\eta'} &f^{0\eta}\cr f^{8\eta'} &f^{8\eta}}\right)
~~=~~\left(\matrix{f_0 \cos\theta_0 &-f_0 \sin\theta_0\cr
f_8 \sin\theta_8 &f_8 \cos\theta_8}\right)
\label{eq:dm}
\ee
Translating from eq.(\ref{eq:dk}), we find
\begin{eqnarray}
f_0 ~=~ 106.6 \pm 4.2 ~{\rm MeV} ~~~~~~~~
f_8 ~=~ 104.8 \pm 1.3 ~{\rm MeV} \nonumber\\
\theta_0 ~=~ -12.3 \pm 3.0 ~{\rm deg} ~~~~~~~~
\theta_8  ~=~ -20.1 \pm 0.7 ~{\rm deg}
\label{eq:dn}
\end{eqnarray}
that is 
\be
{f_0\over f_\pi} ~=~ 1.15 \pm 0.05 ~~~~~~~~
{f_8\over f_\pi} ~=~ 1.13 \pm 0.02
\label{eq:do}
\ee
We emphasise, however, that since our definitions of the decay constants
differ from those in refs.\cite{Leutwyler:1997yr,Kaiser:1998ds}, 
any comparison of the numbers above should be made with care.

The most striking feature of the results (\ref{eq:dj}) for the decay 
constants is how close the diagonal ones, $f^{0\eta'}$ and $f^{8\eta}$,
are to $f_\pi$, even the singlet. Predictably, the off-diagonal ones are 
strongly suppressed, especially $f^{0\eta}$. This supports the approximations 
used in section 2 in deriving the large $N_c$ Witten-Veneziano limit of the 
DGMOR relations.

To see this in more detail, it is interesting to compare numerically the
magnitudes of the various terms appearing in the DGMOR relations, together
with their formal orders in $1/N_c$. We find (all terms in units of 
$10^{-3}{\rm GeV}^4$):
\begin{eqnarray}
&(f^{0\eta'})^2 m_{\eta'}^2 ~[N_c;~9.96] 
+ (f^{0\eta})^2 m_\eta^2 ~[N_c;~0.15] ~~=~~ 
{1\over3} f_\pi^2 m_\pi^2 ~[N_c;~0.06]
+ {2\over3} f_K^2 m_K^2 ~[N_c;~2.07] \nonumber\\
&~~~~~~~~~~~~~~~~~~~~~~~~~~~~~~~+ 6 A ~[1;~7.98] \nonumber\\
\label{eq:ds}\\
\nonumber\\
&f^{0\eta'} f^{8\eta'} m_\eta'^2 ~[N_c;-3.45]
+ f^{0\eta} f^{8\eta} m_{\eta}^2 ~[N_c;~0.67] ~~=~~ 
{2\sqrt2\over3} f_\pi^2 m_\pi^2 ~[N_c;~0.16]
-{2\sqrt2\over3} f_K^2 m_K^2 ~[N_c;-2.94] \nonumber\\
\label{eq:dt}\\
\nonumber\\
&(f^{8\eta'})^2 m_\eta'^2 ~[N_c;~1.19]
+ (f^{8\eta})^2 m_{\eta}^2 ~[N_c;~2.90] ~~=~~ 
-{1\over3}f_\pi^2 m_\pi^2 ~[N_c;-0.06] 
+{4\over3} f_K^2 m_K^2 ~[N_c;~4.15] \nonumber\\
\label{eq:du}\\
\nonumber
\end{eqnarray}

The interesting feature here is the explicit demonstration that the dominant
term in the flavour singlet DGMOR (Witten-Veneziano) relation is the
topological susceptibility factor $6A$, even though it is formally suppressed 
by a power of $1/N_c$ relative to the others.  It is matched by the
subdominant ($O(1)$) contribution to $(f^{0\eta'})^2 m_{\eta'}^2$,
which arises because of the numerically large but $O(1/N_c)$ anomaly-induced
part of the $m_{\eta'}^2$ which survives in the chiral limit.
The numerical results therefore confirm the theoretical intuition
expressed in section 2.

To emphasise this point further, we can summarise the numerical magnitudes
in the combined singlet-octet relation (\ref{eq:bs}), which reduces
to the full Witten-Veneziano formula ({\ref{eq:br}):  
\begin{eqnarray}
(f^{0\eta'})^2 m_{\eta'}^2 ~[N_c;~9.96] + (f^{0\eta})^2 m_\eta^2 ~[N_c;~0.15] 
~~~~~~~~~~~~~~~~~~~~~~~~~~~~~~~~~~~~~~~~~~~~~~~~~~~~\nonumber\\
+~(f^{8\eta'})^2 m_\eta'^2 ~[N_c;~1.19]
+ (f^{8\eta})^2 m_{\eta}^2 ~[N_c;~2.90]
- 2 f_K^2 m_K^2 [N_c;-6.22]
~~=~~ 6A ~[1;~7.98] \nonumber\\
\label{eq:dv}
\end{eqnarray}
The validity  of the large $N_c$ limit leading to eq.(\ref{eq:br}) is 
particularly transparent in this form. Numerically, the surprising accuracy 
of the approximate formula (\ref{eq:br}) is seen to be in part due to a 
cancellation between the underestimates of $f^{8\eta'}$ (taken to be 0) 
and $f_K$ (set equal to $f_\pi$).

Now consider the decay formulae themselves. The numerical
magnitudes and $1/N_c$ orders of the various contributions in this case are
(in units of $10^{-3})$:
\be
f^{0\eta'} g_{\eta'\c\c} ~[N_c;~3.23]
+ f^{0\eta} g_{\eta\c\c} ~[N_c;~0.57]
+ {\sqrt6} A g_{G\c\c} ~[1;~-0.005 \pm 0.23] 
=~~a_{\rm em}^0 {\a_{\rm em}\over\pi} ~[N_c;~3.79]
\label{eq:dw}
\ee
and
\be
f^{8\eta'}g_{\eta'\c\c} ~[N_c;-1.12]
+ f^{8\eta} g_{\eta\c\c} ~[N_c;~2.46] 
~~=~~a_{\rm em}^8 {\a_{\rm em}\over\pi} ~[N_c;~1.34]
\label{eq:dx}
\ee

The interest here is in the realisation of the $1/N_c$ approximation
in the flavour singlet decay formula. As explained in section 2,
the coupling $g_{G\c\c}$ is renormalisation group invariant\footnote{The 
topological susceptibility parameter $A$ is also renormalisation group 
invariant.} and $O(1/N_c)$ suppressed
and our conjecture is that such terms would indeed be relatively small.
Remarkably, evaluated at the central value of the topological 
susceptibility found in ref.\cite{DelDebbio:2004ns}, the coupling
$g_{G\c\c}$ is essentially zero. This is probably a numerical coincidence, 
since we can not think of a dynamical reason why this coupling should 
vanish identically. What is more reasonable is to consider its value
across the range of error of the topological susceptibility. In this
case, we see from eq.(\ref{eq:dw}) that the suppression is numerically 
still under $10\%$, which is closer to that expected for a typical OZI 
correction although still remarkably small.

This is a very encouraging result. First, it increases our confidence
that we are able to identify quantities where the OZI, or leading $1/N_c$,
approximation is likely to be numerically good. It also shows that
$g_{G\c\c}$ gives a contribution to the decay formula which is entirely
consistent with its picturesque interpretation as the coupling of the
photons to the anomaly-induced gluonic component of the $\eta'$.
{\it A posteriori}, the fact that its contribution is at most $10\%$
explains the general success of previous theoretically inconsistent
phenomenological parametrisations of $\eta'$ decays in which the
naive current algebra formulae omitting the gluonic term are used.

\section{Further discussion}

We now summarise our results so far and discuss a number of aspects of their
validity and applicability to other processes, both in low-energy
pseudoscalar meson physics and in high-energy processes such as
deep inelastic scattering or photoproduction.

In this paper, we have seen how the uncontroversial radiative decay
formula
\be
f^{8\eta'}g_{\eta'\c\c} + f^{8\eta} g_{\eta\c\c} 
~~=~~{1\over\sqrt3} {\a_{\rm em}\over\pi}
\label{eq:ea}
\ee
together with the three DGMOR relations
\begin{eqnarray}
&(f^{0\eta'})^2 m_{\eta'}^2 + (f^{0\eta})^2 m_\eta^2 ~~=~~ 
{1\over3} \bigl(f_\pi^2 m_\pi^2 + 2 f_K^2 m_K^2\bigr) + 6 A 
\label{eq:eb}\\
\nonumber\\
&f^{0\eta'} f^{8\eta'} m_\eta'^2 + f^{0\eta} f^{8\eta} m_{\eta}^2 ~~=~~ 
{2\sqrt2\over3}\bigl(f_\pi^2 m_\pi^2 - f_K^2 m_K^2\bigr)
\label{eq:ec}\\
\nonumber\\
&(f^{8\eta'})^2 m_\eta'^2 + (f^{8\eta})^2 m_{\eta}^2 ~~=~~ 
-{1\over3}\bigl(f_\pi^2 m_\pi^2 - 4 f_K^2 m_K^2\bigr)
\label{eq:ed}\\
\nonumber
\end{eqnarray}
lead to the following identification of the four pseudoscalar meson
decay constants in the mixed $\eta',\eta$ sector:
\be
\left(\matrix{f^{0\eta'} &f^{0\eta}\cr f^{8\eta'} &f^{8\eta}}\right)
~~=~~\left(\matrix{104.2 \pm 4.0 & 22.8 \pm 5.7  \cr
-36.1 \pm 1.2 & 98.4 \pm 1.4 }\right)~{\rm MeV}
\label{eq:ee}
\ee

The novel feature is the use of the flavour singlet DGMOR relation.
We emphasise again that this is, within the framework of PCAC or chiral
Lagrangians, an exact result in the sense of being entirely independent 
of the $1/N_c$ expansion. Its large-$N_c$ limit is the well-established
Witten-Veneziano formula. This determination of the decay constants is
therefore on firm theoretical ground and provides a sound basis for
phenomenology.\footnote{Note that the PDG tables currently 
do not quote values for the flavour singlet decay constants because of 
the subtleties in their definition. A good case can therefore be made
for adopting the definitions and experimental numbers presented here 
as a simple and theoretically well-motivated parametrisation of the data.}

The use of the flavour singlet DGMOR formula relies 
on the input of the non-perturbative parameter $A$ which controls 
the topological susceptibility in QCD. In time, lattice calculations
should be able to determine this number accurately from simulations
in full QCD with massive, dynamical quarks. For the moment, we have to
rely on the result obtained from the pure Yang-Mills calculation,
which we have argued using $1/N_c$ ideas should be a good approximation.
This temporary approximation is the only place where $1/N_c$ enters
the determination of the decay constants.

To illustrate the dependence of the decay constants on the topological
susceptibility, we have plotted the singlet decay constants 
$f^{0\eta'}$ and $f^{0\eta}$ against $A$ in Fig.~1. It is clear from 
the DGMOR and radiative decay relations that the octet decay constants 
$f^{8\eta'}$ and $f^{8\eta}$ are themselves independent of $A$.

\FIGURE
{\epsfxsize=7cm\epsfbox{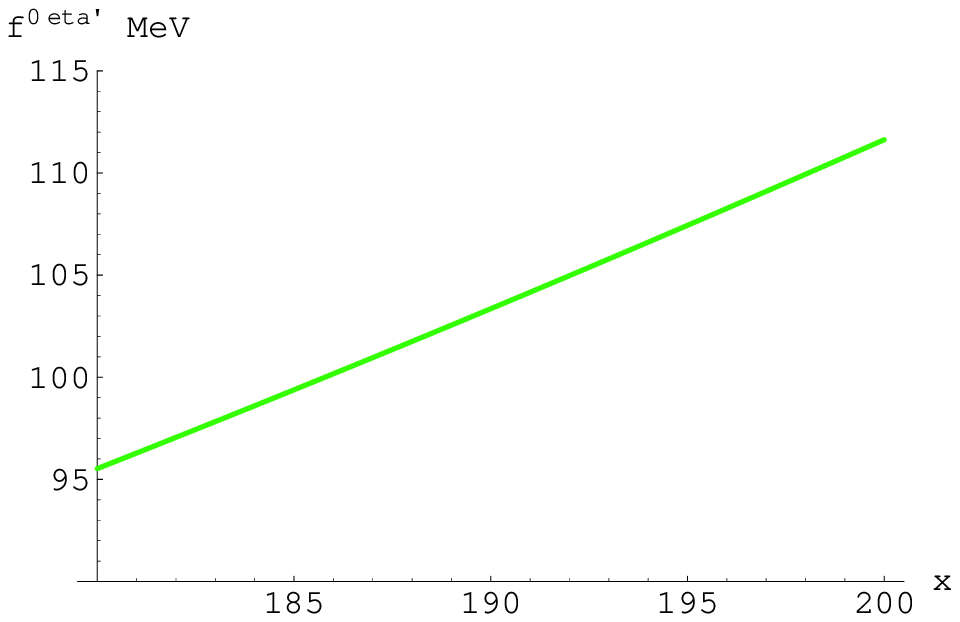}\hskip1cm
\epsfxsize=7cm\epsfbox{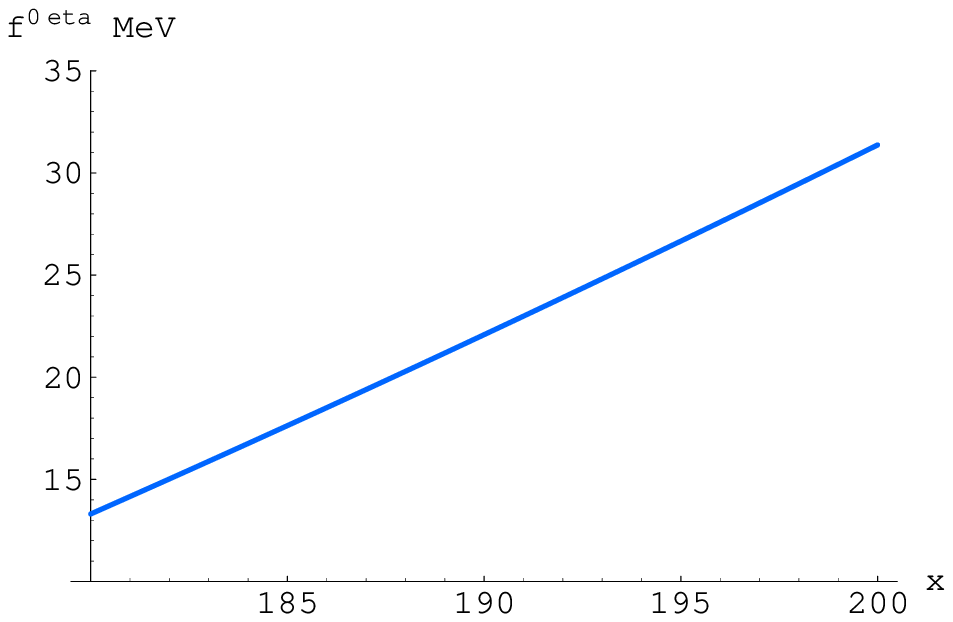}
\caption{The decay constants $f^{0\eta'}$ and $f^{0\eta}$ as functions of 
the non-perturbative parameter $A = (x~{\rm MeV})^4$ which determines the
topological susceptibility in QCD.}
}

The remaining part of our analysis is the flavour singlet decay formula
\be
f^{0\eta'} g_{\eta'\c\c} + f^{0\eta} g_{\eta\c\c}f^{0\eta} g_{\eta\c\c} + 
{\sqrt6} A g_{G\c\c} 
~~=~~{2\sqrt2\over\sqrt3} {\a_{\rm em}\over\pi}
\label{eq:ef}
\ee
Here, because the $\eta'$ is not a NG boson even in the chiral limit,
the naive PCAC formula acquires an extra term.  In our formulation,
this is the parameter $g_{G\c\c}$ which we have argued may reasonably,
though certainly non-rigorously, be interpreted as the coupling of the
photons to the anomaly-induced gluonic component of the $\eta'$, i.e.~the
component which removes its NG boson status.  The picture is simple
and attractive. The experimental value of the unphysical coupling, 
$g_{G\c\c} = -0.001 \pm 0.072 {\rm GeV}^{-4}$, 
means that its contribution to the decay formula is under $10\%$. 
This is consistent with our expectations for a renormalisation group 
invariant, $1/N_c$ (or OZI) suppressed quantity.

This is illustrated in Fig.~2, where we have shown the size of the 
contributions of the various terms in eq.(\ref{eq:ef}), showing explicitly
their dependence on the topological susceptibility.

\FIGURE
{\epsfxsize=9cm\epsfbox{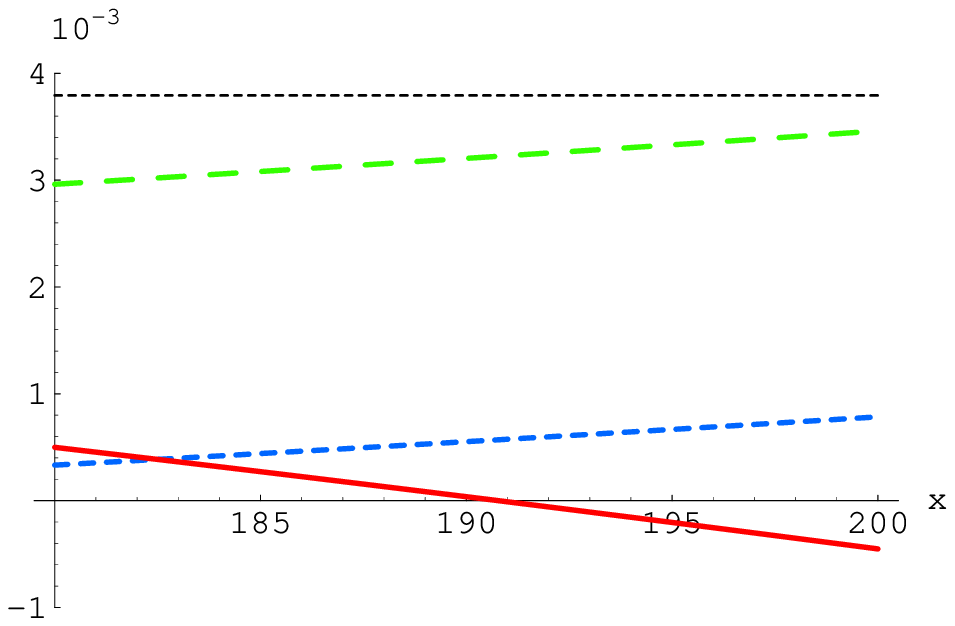}
\caption{This shows the relative sizes of the contributions to the flavour 
singlet radiative decay formula (\ref{eq:ef}) expressed as functions of the
topological susceptibility parameter $A = (x~{\rm MeV})^4$.  The dotted 
(black) line denotes ${2\sqrt2\over\sqrt3} {\a_{\rm em}\over\pi}$. The
dominant contribution comes from the term $f^{0\eta'} g_{\eta'\c\c}$, denoted
by the long-dashed (green) line, while the short-dashed (blue) line denotes
$f^{0\eta} g_{\eta\c\c}$. The contribution from the gluonic coupling,
${\sqrt6} A g_{G\c\c}$, is shown by the solid (red) line.}
}

However, while the flavour singlet decay formula is sensible and theoretically
consistent, it is necessarily non-predictive. To be genuinely
useful, we would need to find another process in which the same coupling
enters. The problem here is that, unlike the decay constants which are
universal, the coupling $g_{G\c\c}$ is process-specific just like
$g_{\eta'\c\c}$ or $g_{\eta\c\c}$. There are of course many other 
processes to which our methods may be applied such as $\eta'(\eta)\rta V\c$,
where $V$ is a flavour singlet vector meson $\rho,\omega,\phi$, or $\eta'
(\eta)\rta \pi^+ \pi^- \c$. The required flavour singlet formulae may readily
be written down, generalising the naive PCAC formulae. However, each will 
introduce its own gluonic coupling, such as $g_{GV\c}$. Although strict
predictivity is lost, our experience with the two-photon decays suggests that
these extra couplings will give relatively small, at most $O(10-20\%$), 
contributions if like $g_{G\c\c}$ they can be identified as RG invariant 
and $1/N_c$ suppressed.\footnote{Notice though that this must be used 
carefully. For example, the coupling $g_{GNN}$ which enters the $U(1)_A$ 
Goldberger-Treiman relation alongside the $\eta'$-nucleon coupling 
$g_{\eta' NN}$ is $1/N_c$ suppressed but {\it not} RG invariant and is 
expected to display large deviations from its large $N_c$ or OZI limit. 
See section 6.} 
This observation restores at least a reasonable degree of predictivity 
to the use of PCAC methods in the $U(1)_A$ sector.

Interestingly, these novel gluonic couplings may also arise in high-energy 
processes. For example, the standard analysis of the two-photon 
deep-inelastic scattering process $e^+ e^- \rta e^+ e^- X$ reduces the 
problem of finding the first moment of the polarised photon structure
function $g_1^\c$ to a non-perturbative evaluation of the off-shell
matrix element $\la\c|J_{\m 5}^0|\c\ra$. The difference with 
$\eta'(\eta)\rta\c\c$ is that in the DIS scenario, the photons are
off-shell and the interest in the first moment sum rule for $g_1^\c$
is precisely how it depends on the target photon momentum 
\cite{Narison:1992fd, Shore:1992pm, Shore:2004cb}.
Nevertheless, the problem may be formulated in terms of form factors
$g_{\eta'(\eta)\c\c}(k^2)$ and $g_{G\c\c}(k^2)$ of which the
couplings discussed here are simply the $k^2=0$ limit \cite{Shore:2004cb}. 
It has also been suggested that the gluonic couplings $g_{G\phi\c}$ and 
$g_{GNN}$ could play a dominant role in the photoproduction process 
$\c N \rta \phi N$ \cite{Kochelev:2001nv}. This interpretation is less clear, 
but it is an interesting subject for future work to look at a variety of 
electro or photoproduction experiments in the light of the PCAC methods 
developed here.

\section{Pseudoscalar meson couplings of the nucleon and the 
$U(1)_A$ Goldberger-Treiman relation}

A further particularly interesting application of these ideas is to the
pseudoscalar couplings of the nucleon. For the pion, the relation
between the axial-vector form factor of the nucleon and the pion-nucleon
coupling $g_{\pi NN}$ is the well-known Goldberger-Treiman relation. 
Here, we are concerned with its generalisation to the flavour-singlet 
sector, which involves the anomaly and gluon topology. This $U(1)_A$ 
Goldberger-Treiman relation was first developed in refs.\cite{Veneziano:1989ei,
Shore:1990zu, Shore:1991dv}.
In this case, the corresponding high-energy process involves the measurement
of the first moment of the polarised structure function of the nucleon
$g_1^N$ in deep-inelastic scattering. In the flavour-singlet sector, this
is the so-called `proton spin' problem. 
(For reviews, see e.g.~refs\cite{Shore:1998dn, Bass:2004xa}.)

The axial-vector form factors are defined from
\be
\langle N|J_{\mu 5}^a|N\rangle ~=~
2m_N \Bigl( G_A^a(k^2) s_\mu  +  G_P^a(k^2) k.s k_\mu \Bigr)
\label{eq:fa}
\ee
where $s_\mu = \bar u \c_\mu \c_5 u /2m_N$ is the covariant spin vector.
In the absence of a massless pseudoscalar, only the form factors $G_A(0)$
contribute at zero momentum.  
Using the `$U(1)_A$ PCAC' substitution (\ref{eq:cm}) for 
$\partial^\mu J_{\mu 5}^a$ and repeating the steps explained in section 3
(particularly at eq.(\ref{eq:co})), we straightforwardly find the
following generalisation of the Goldberger-Treiman relation:
\be
2m_N G_A^a(0) ~~=~~ f^{a\a} g_{\eta^\a NN} ~+~ \sqrt{2n_f} A g_{GNN}\d_{a0}
\label{eq:fb}
\ee
with the obvious definition of the gluonic coupling $g_{GNN}$ in analogy
to $g_{\eta^\a NN}$.  

For the individual flavour components, this 
reads (abbreviating $G_A^a(0) = G_A^a$):
\begin{eqnarray}
2m_N G_A^3 ~~=~~ f_\pi g_{\pi NN}~~~~~~~~~~~~~~~~~~~~&
\label{eq:fc} \\
\nonumber\\
2m_N G_A^8 ~~=~~ f^{8\eta'} g_{\eta' NN} ~+~ f^{8\eta} g_{\eta NN}&
\label{eq:fd} \\
\nonumber\\
2m_N G_A^0 ~~=~~ f^{0\eta'} g_{\eta' NN} ~+~ f^{0\eta} g_{\eta NN}&
~+~ \sqrt{6} A g_{GNN}
\label{eq:fe} 
\end{eqnarray}
Eq.(\ref{eq:fe}) is the $U(1)_A$ Goldberger-Treiman relation.\footnote{The 
original form as quoted in ref.\cite{Shore:1991dv} applies to the chiral limit 
and reads, in the notation of \cite{Shore:1991dv} but allowing for our 
different normalisation of the singlet,
$$
2m_N G_A^0 ~=~ F g_{\eta' NN} + {1\over\sqrt{2n_f}} F^2 m_{\eta'}^2 g_{GNN}
$$
where $F$ is a RG invariant decay constant defined from the two-point
Green function of the pseudoscalar field $\phi_5$. In the chiral limit,
where there is no $SU(3)$ mixing, this is reproduced by the definition
(\ref{eq:cl}) of $f^{0\eta'}$. The off-diagonal decay constant 
$f^{0\eta}$ vanishes. The final term is reproduced by eq.(\ref{eq:fe})
by virtue of the flavour-singlet DGMOR relation (\ref{eq:bg}) in the chiral
limit.} Notice that the flavour-singlet coupling $G_A^0$ is not renormalisation
group invariant and so depends on the RG scale. This is reflected in the
RG non-invariance of the gluonic coupling $g_{GNN}$ \cite{Shore:1991dv}.

In the notation that has become standard in the DIS literature, the
axial couplings are
\be
G_A^3 ~=~ {1\over2}~ a^3 ~~~~~~~~
G_A^8 ~=~ {1\over 2\sqrt{3}}~ a^8 ~~~~~~~~
G_A^0 ~=~ {1\over\sqrt{6}}~ a^0
\label{eq:ff}
\ee
and have the following interpretation in terms of parton distribution
functions:
\begin{eqnarray}
a^3 ~=~ \D u - \D d ~~~~~~~~~~~~~~~~~~~&\nonumber \\
a^8 ~=~ \D u + \D d - 2\D s ~~~~~~~~~~&\nonumber \\
a^0 ~=~ \D u + \D d + \D s - {3\a\over2\pi} \D g &
\label{eq:fg}
\end{eqnarray}
Experimentally,
\be 
a^3 ~=~ 1.267 \pm 0.004 ~~~~~~~~
a^8 ~=~ 0.585 \pm 0.025
\label{eq:fh}
\ee
from low-energy data, while the latest result for $a^0$ quoted by the
COMPASS collaboration \cite{Mallot,Ageev:2005gh} is
\be
a^0|_{Q^2=4{\rm GeV}^2} ~=~ 0.237{}^{+0.024}_{-0.029}
\label{eq:fi}
\ee
It is the fact that $a^0$ is much less than $a^8$, as would be predicted 
on the basis of the simple quark model (the Ellis-Jaffe sum rule 
\cite{Ellis:1973kp}),
that is known as the `proton spin' problem. For a careful analysis of
the distinction between the angular momentum (spin) of the proton and the 
axial coupling $a^0$, see however refs.\cite{Shore:1999be,Shore:2000ca}.

From the standard Goldberger-Treiman relation (\ref{eq:fc}), we immediately
find the following result for the (dimensionless) pion-nucleon coupling,
\be
g_{\pi NN} ~=~ 12.86 \pm 0.06
\label{eq:fj}
\ee
consistent to within $\sim 5\%$ with the experimental value $13.65 (13.80)
\pm 0.12$ (depending on the dataset used) \cite{Bugg:2004cm}.

In an ideal world where $g_{\eta NN}$ and $g_{\eta' NN}$ were both known,
we would now verify the octet formula (\ref{eq:fd}) then determine
the gluonic coupling $g_{GNN}$ from the singlet Goldberger-Treiman
relation (\ref{eq:fe}). However, the experimental situation with the 
$\eta$ and $\eta'$-nucleon couplings is far less clear. 
(See refs.\cite{Moskal:2004cm, Bass:2001ix}
for reviews of the relevant experimental literature and recent results.)
One would hope to determine these couplings from the near threshold 
production of the $\eta$ and $\eta'$ in nucleon-nucleon collisions, i.e. 
$pp\rightarrow pp\eta$ and $pp\rightarrow pp\eta'$, measured for example
at COSY-II \cite{Moskal:2004nw}. However, the $\eta$ production is 
dominated by the $S_{11}$ nucleon resonance $N^*(1535)$ which decays
to $N\eta$, and as a result very little is known about $g_{\eta NN}$ itself.
The detailed production mechanism of the $\eta'$ is not well understood.
However, since there is no known baryonic resonance decaying into $N\eta'$, 
we may simply assume that the reaction $pp\rightarrow pp\eta'$ is 
driven by the direct coupling supplemented by heavy-meson exchange. This 
allows an upper bound to be placed on $g_{\eta' NN}$ and on this basis 
ref.\cite{Moskal:1998pc} quotes $g_{\eta' NN}< 2.5$.  This is supported by
an analysis \cite{Nakayama:2005ts} of very recent data from CLAS 
\cite{Dugger:2005du}} on the photoproduction reaction 
$\c p \rightarrow p \eta'$. Describing the cross-section data with a model 
comprising the direct coupling together with $t$-channel meson exchange 
and $s$ and $u$-channel resonances, it is found that equally good fits can 
be obtained for several values of $g_{\eta' NN}$ covering the whole region 
$0 < g_{\eta' NN} < 2.5$.  

In view of this experimental uncertainty, we shall use the octet and singlet 
GT relations to plot the predictions for $g_{\eta NN}$ and $g_{GNN}$ as a 
function of the $\eta'$-nucleon coupling in the range $0< g_{\eta' NN}< 2.5$.
The results are shown in Fig.~3.

\FIGURE
{\epsfxsize=7cm\epsfbox{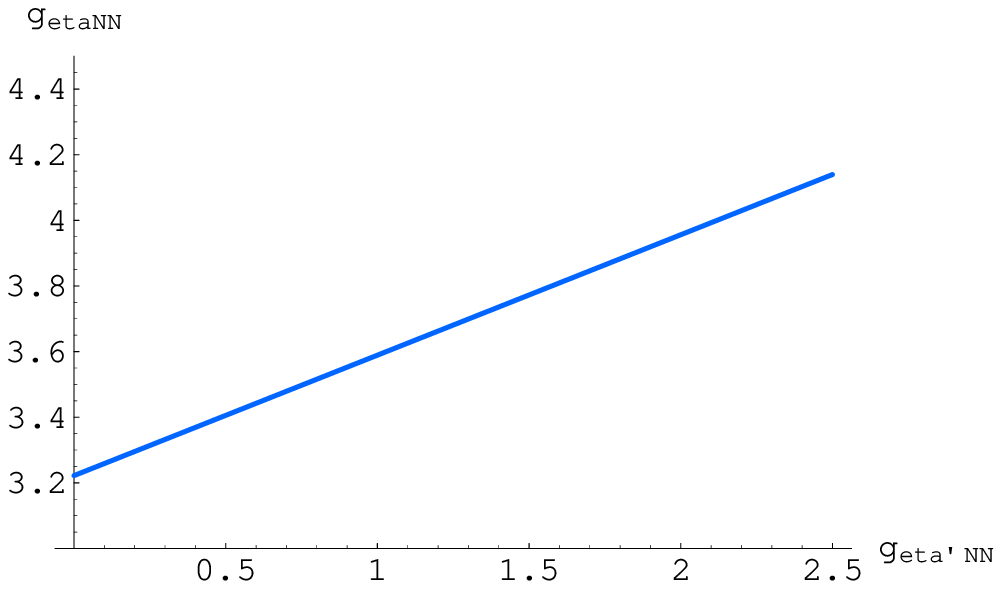}\hskip1cm
\epsfxsize=7cm\epsfbox{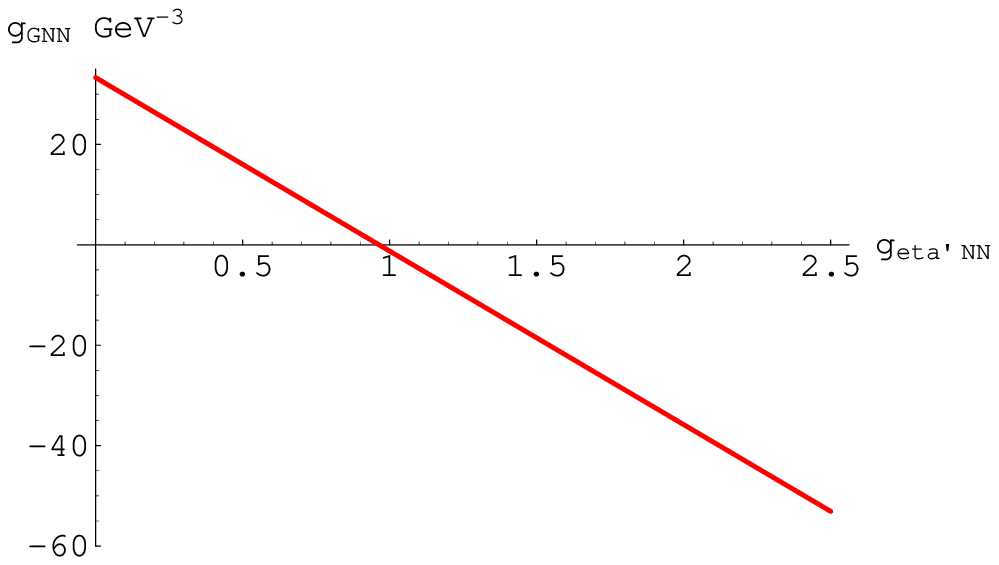}
\caption{These figures show the dimensionless $\eta$-nucleon coupling 
$g_{\eta NN}$ and the gluonic coupling $g_{GNN}$ in units of 
${\rm GeV}^{-3}$ expressed as functions of the experimentally uncertain 
$\eta'$-nucleon coupling $g_{\eta' NN}$, as determined from the flavour octet 
and singlet Goldberger-Treiman relations (\ref{eq:fd}) and (\ref{eq:fe}).}
}

Our main interest in the $U(1)_A$ Goldberger-Treiman relation lies of course
in the gluonic coupling $g_{GNN}$. Unlike its counterpart $g_{G\c\c}$ in
the radiative decay formula, $g_{GNN}$ is not a renormalisation group
invariant coupling. However, like $g_{G\c\c}$ it is suppressed at large $N_c$.
The various terms in eq.(\ref{eq:fe}) have the following orders:
$G_A = O(N_c)$, ~$f^{0\eta}, f^{0\eta'} = O(\sqrt{N_c})$, ~$A = O(1)$,
~$g_{\eta NN}, g_{\eta' NN} = O(\sqrt{N_c})$, ~$g_{GNN} = O(1)$. 
So the final term $A g_{GNN}$ is $O(1)$, down by a power of $1/N_c$ compared 
to all the others, which are $O(N_c)$.

The intuition we have developed through experience with flavour singlet physics
and the large-$N_c$ expansion is that while we expect $O(1/N_c)$ suppressed
RG invariant quantities to be numerically small, in line with expectations
from the OZI rule, we do not expect this to be necessarily true for
RG non-invariant quantities such as $g_{GNN}$.\footnote{More precisely,
we expect the OZI approximation to be unreliable for quantities which have
a different RG behaviour in QCD itself and in the OZI limit. The complicated
RG non-invariance of $g_{GNN}$ in QCD is induced by the axial anomaly, since
$G_A^0$ itself is required to scale with the anomalous dimension $\c$
of the flavour-singlet current $J_{\mu 5}^0$ and all the other terms in
the $U(1)_A$ GT relation are RG invariant. The anomaly, and thus the 
anomalous dimension $\c$, vanishes in the large-$N_c$ limit leaving
$g_{GNN}$ RG invariant as $N_c\rightarrow \infty$.} 
So unlike $g_{G\c\c}$ in the flavour-singlet radiative decay formula, 
we would not be surprised if $g_{GNN}$ makes a sizeable numerical contribution 
to the $U(1)_A$ GT relation.\footnote{Of course, since $g_{GNN}$ is defined
with dimension ${\rm GeV}^{-3}$ whereas $g_{\eta NN}$ and $g_{\eta' NN}$ are 
dimensionless, we cannot make a direct comparison of the couplings themselves.}

This is quantified in Fig.~4, where we have plotted the contribution of each
term in the $U(1)_A$ GT relation as a function of $g_{\eta' NN}$. We see
immediately that the contribution from $f^{0\eta}g_{\eta NN}$ is relatively
constant around $0.08$, compared with $2m_N G_A^0 \sim 0.18$. This means
that the variation of $f^{0\eta'}g_{\eta' NN}$ over the experimentally 
allowed range is compensated entirely by the variation of $\sqrt{6}A g_{GNN}$.
For generic values of $g_{\eta' NN}$, there is no sign of a significant
suppression of the contribution of the gluonic coupling $g_{GNN}$ relative
to the others. This should be contrasted with the corresponding plot for
$g_{G\c\c}$ in the radiative decay formula (Fig.~2).

\FIGURE
{\epsfxsize=9cm\epsfbox{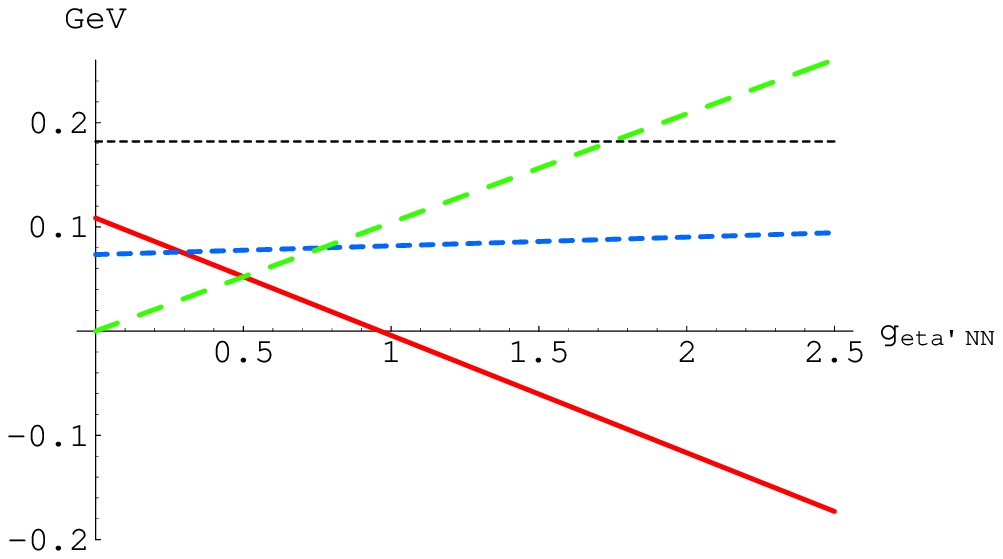}
\caption{This shows the relative sizes of the contributions to the
$U(1)_A$ Goldberger-Treiman relation from the individual terms in 
eq.(\ref{eq:fe}), expressed as functions of the coupling $g_{\eta' NN}$. 
The dotted (black) line denotes $2m_N G_A^0$. The long-dashed (green) line 
is $f^{0\eta'}g_{\eta' NN}$ and the short-dashed (blue) line is 
$f^{0\eta}g_{\eta NN}$. The solid (red) line shows the contribution of
the novel gluonic coupling, $\sqrt{6}Ag_{GNN}$, where $A$ determines the
QCD topological susceptibility.}
}

To see this in more detail, consider a representative value, 
$g_{\eta' NN}=2.0$, which would correspond to the direct coupling 
contributing substantially to the cross sections for $pp\rightarrow pp\eta'$ 
and $\c p \rightarrow p\eta'$. In this case, $g_{\eta NN} = 3.96 \pm 0.16$ 
and $g_{GNN} = -14.6 \pm 4.3 {\rm GeV}^{-3}$.
The contributions to the $U(1)_A$ GT relation are then (in GeV):
\be
2m_N G_A^0 [N_c;~0.18] ~~=~~ 
f^{0\eta'} g_{\eta' NN} [N_c;~0.21] ~+~ 
f^{0\eta} g_{\eta NN} [N_c;~0.09] ~+~ 
\sqrt{6} A g_{GNN} [O(1);~-0.12]
\label{eq:fk}
\ee
The anomalously small value of $G_A^0$ compared to $G_A^8$ is due to the 
partial cancellation of the sum of the two meson-coupling terms, which 
together contribute close to the expected OZI value ($2m_N G_A^8 = 0.32$), 
by the gluonic coupling $g_{GNN}$. Although this is formally $O(1/N_c)$ 
suppressed, numerically it gives the dominant contribution to the large OZI 
violation in $G_A^0$. This is in line with our expectations and would provide 
further evidence that the insights developed in our body of work on both 
low-energy $\eta$ and $\eta'$ physics and related high-energy phenomena 
such as the `proton spin' problem are on the right track.

However, it may still be that $g_{\eta' NN}$ is significantly smaller,
implying a relatively small contribution to the production reactions
$pp\rightarrow pp\eta$ and $\c p \rightarrow p\eta'$ from the direct coupling.
In particular, there is a range of $g_{\eta' NN}$ around $0.7 - 1.3$ where  
the gluonic coupling $g_{GNN}$ only contributes to the $U(1)_A$ GT relation
at the level expected of a typical OZI-suppressed quantity, despite its
RG non-invariance. In the extreme case where $g_{\eta' NN}=1.0$, we have 
$g_{\eta NN}= 3.59 \pm 0.15$,~ $g_{GNN} = 0.5 \pm 3.8 {\rm GeV}^{-3}$,  
and the contributions to the $U(1)_A$ GT formula become (in {\rm GeV}):
\be
2m_N G_A^0 [N_c;~0.18] ~~=~~ 
f^{0\eta'} g_{\eta' NN} [N_c;~0.10] ~+~ 
f^{0\eta} g_{\eta NN} [N_c;~0.08] ~+~ 
\sqrt{6} A g_{GNN} [O(1);~-0.004]
\label{eq:fl}
\ee
This scenario would be similar to the radiative decays, where we
found that the corresponding coupling $g_{G\c\c} \simeq 0$ using the central
value of the lattice determination of the topological susceptibility and 
contributes only at $O(10\%)$ within the error bounds on $A$. This would then
suggest that RG non-invariance is not critical after all and the $O(1/N_c)$
suppressed gluonic couplings are indeed numerically small. It would also leave
open the possibility that all couplings of type $g_{GXX}$ are close to zero,
which in the picturesque interpretation discussed earlier would imply that
the gluonic component of the $\eta'$ wave function is small. The suppression
of $G_A^0$ relative to $G_A^8$ would then not be due to gluonic, 
anomaly-induced OZI violations but rather to the particular nature of the
flavour octet-singlet mixing in the $\eta - \eta'$ sector. In the parton 
picture of the `proton spin' problem, this would be a hint that the 
suppression in $a^0$ is not primarily due to the polarised gluon distribution
$\D g$ but also involves a strong contribution from the polarised
strange quark distribution $\D s$.

Although we would consider this alternative scenario rather surprising and 
prefer the more theoretically motivated interpretation of the flavour singlet
sector in which $g_{\eta' NN} \simeq 2$ and gluon topology plays an important 
role, ultimately the decision rests with experiment. Clearly, a reliable 
determination of $g_{\eta' NN}$, or equivalently $g_{\eta NN}$, would shed 
considerable light on the $U(1)_A$ dynamics of QCD.

\vskip1.5cm

\acknowledgments

I would like to thank G.~Veneziano for interesting comments and collaboration
on the original investigations of $U(1)_A$ physics on which this paper is based.
This work is supported in part by PPARC grants PPA/G/O/2002/00470 and 
PP/D507407/1.

\vfill\eject

\appendix

\section{Appendix:~Comparison with chiral Lagrangians}

It is useful to compare the results presented in this paper with those
arising in extensions of the chiral Lagrangian formalism to include the
$\eta'$ and low-energy flavour singlet physics. The large-$N_c$ expansion
plays a crucial role in this approach, since it is only in the limit
$N_c \rta \infty$ that the anomaly disappears, the chiral symmetry is
enlarged to $U(3)_L \times U(3)_R$ and the $\eta'$ appears as a light
NG degree of freedom to be included in the fundamental fields 
$\varphi^a$ $(a=0,1 \ldots 8)$ of the chiral Lagrangian. Large-$N_c$ chiral 
Lagrangians have been developed by a number of authors, 
notably ref.\cite{Herrera-Siklody:1996pm}, though here we 
shall focus on the results obtained by Kaiser and Leutwyler 
\cite{Leutwyler:1997yr,Kaiser:1998ds,Kaiser:2000gs}.

The elegance of chiral Lagrangians should not obscure the fact 
that all the dynamical approximations we have made in deriving our results,
such as pole dominance by NG bosons, weak momentum dependence of pole-free 
quantities like decay constants and couplings and a judicious use of 
the $1/N_c$ expansion, are necessarily also made in the chiral Lagrangian 
formalism, where they are built in to the structure of the initial 
Lagrangian. Indeed, our results can also be systematised into an
effective Lagrangian (see ref.\cite{Shore:1999tw} for details). 
The principal merit of the chiral/effective Lagrangian approach in general
is in providing a systematic way of calculating higher-order corrections.

The physical results obtained using the two methods should therefore
be equivalent. However, a number of our definitions, notably of the
flavour singlet decay constants, differ from those made by Kaiser and Leutwyler
so the comparison is not straightforward. However, we would argue that in many
ways the formalism developed in this paper provides a better starting point
for the description of $U(1)_A$ phenomenology, especially in
the use of RG invariant decay constants and our natural generalisation of the 
Witten-Veneziano formula as the flavour singlet DGMOR relation.

The fundamental fields in the KL chiral Lagrangian are assembled into matrices
$U= \exp[i\varphi^a T^a]$ so that, up to mixing, the fields are in one-to-one 
correspondence with the NG bosons in the chiral and large-$N_c$ 
limits.\footnote{We use the same normalisation for the generators as in the 
rest of this paper, so our singlet field differs from the $\psi$ of 
refs.\cite{Leutwyler:1997yr,Kaiser:1998ds,Kaiser:2000gs} by $\varphi^0 = 
\sqrt{2\over3}~\psi$. The normalisation of the singlet currents and decay 
constants is, however, the same. We assume isospin symmetry and represent the
quark mass matrix by $M = {\rm diag}(m_u,m_d,m_s)$ with $m_u=m_d=m$.} 
The Lagrangian is a simultaneous 
expansion in three parameters - momentum ($p$), quark mass ($m$) and $1/N_c$. 
For bookkeeping purposes, KL consider these to be related as follows:
$p^2 = O(\d)$, $m = O(\d)$, $1/N_c = O(\d)$, and expand consistently in the
small parameter $\d$. It will be clear, however, that this is mere
bookkeeping and should be treated with considerable caution. As we have
seen, the realisation of the $1/N_c$ expansion in the singlet sector is
extremely delicate and it cannot simply be assumed that quantities,
especially RG non-invariant ones, that are $1/N_c$ suppressed are 
necessarily numerically small or indeed of $O(p^2,m)$.

Nevertheless, arranging the allowed terms in the chiral Lagrangian
according to their order in $\d$, KL find:
\be
{\cal L} ~~=~~ {\cal L}_0 ~+~ {\cal L}_1 ~+~ {\cal L}_2 ~+~ \ldots
\ee
where 
\be
{\cal L}_0 ~~=~~ {1\over4} F^2 {\rm tr}(\pl_\m U^\dagger \pl^\m U)
+ {1\over2}F^2 B  {\rm tr}(M^\dagger U + U^\dagger M)
- {3\over4} \t (\varphi^0)^2
\label{eq:ga}
\ee
and
\begin{eqnarray}
&{\cal L}_1 ~~=~~ 
2B L_5 {\rm tr}\Bigl(\pl_\m U^\dagger\pl^\m U(M^\dagger U + U^\dagger M)\Bigr)
~+~ 4B^2 L_8 {\rm tr}\Bigl((M^\dagger U)^2 + (U^\dagger M)^2\Bigr)
~~~~~~~~\nonumber \\
&~~~~+~ {1\over8} F^2 \L_1 \pl^\m \varphi^0 \pl_\m \varphi^0
~+~ {i\over6}\sqrt{3\over2} F^2 B \L_2 \varphi^0  
{\rm tr}(M^\dagger U - U^\dagger M)
\label{eq:gb}
\end{eqnarray}
Most of the physics we are interested in can be derived from ${\cal L}_0 + 
{\cal L}_1$. However, we also encounter some of the terms in the next order,
\begin{eqnarray}
&{\cal L}_2 ~~=~~ 2B L_4 {\rm tr}(\pl_\m U^\dagger \pl^\m U)
{\rm tr}(M^\dagger U + U^\dagger M)
~~~~~~~~~~~~~~~~~~~~~~~~~~~~~~~~~~~~~~~~~~~~~~~~~~~~~~~~~~~~~~~~~~~ 
\nonumber \\
&+~ 4B^2 L_6 \Bigl({\rm tr}(M^\dagger U + U^\dagger M)\Bigr)^2 
~+~ 4B^2 L_7 \Bigl({\rm tr}(M^\dagger U - U^\dagger M)\Bigr)^2 
~~~~~~~~~~~~~~~~~~~~~~~~~~~~~\nonumber \\
&-~ 2i \sqrt{3\over2}  B L_{18} \pl_\m \varphi^0 {\rm tr}(M^\dagger \pl^\m U 
- \pl^\m U^\dagger M)
~-~ 4i \sqrt{3\over2}  B^2 L_{25} \varphi^0 {\rm tr}\Bigl((M^\dagger U)^2 + 
(U^\dagger M)^2\Bigr) \nonumber \\
&~+~\ldots~~~~~~~~~~~~~~~~~~~~~~~~~~~~~~~~~~~~~~~~~~~~~~~~~~~~~~~~~~~~~~
~~~~~~~~~~~~~~~~~~~~~~~~~~~~~~~~~~~~~~~
\label{eq:gc}
\end{eqnarray}
Here, $M$ is the quark mass matrix, $B$ sets the scale of the quark condensate,
$F$ is the leading-order decay constant before $SU(3)$ breaking, and $\t$ is
a parameter which is identified at leading order in $1/N_c$ with the 
topological susceptibility of pure Yang-Mills.  Their $1/N_c$ orders are:~ 
$F^2 = O(N_c)$ and $B,\t = O(1)$.
The coefficients entering at higher order have the following dependence:~
$\L_1, \L_2 = O(1/N_c)$,~ $L_5, L_8 = O(N_c)$ and
$L_4, L_6, L_7, L_{18}, L_{25} = O(1)$.

KL define their decay constants $F^a_{P}$ ($P$ = $\pi$, $K$, $\eta$, $\eta'$)
in the conventional way as the couplings to the axial-vector currents:
\be
\langle0|J_{\mu 5}^a|P\rangle ~=~ i k_\m F^a_P
\label{eq:gd}
\ee
As a consequence, the singlet decay constants $F^0_{\eta'}$ and $F^0_{\eta}$
are not RG invariant but scale with the usual anomalous dimension $\c$
corresponding to the multiplicative renormalisation of the singlet current.
It follows that the parameters $(1+\L_1)$ and $\t$ are also not RG invariant,
scaling with anomalous dimension $2\c$.\footnote{Explicitly, the parameters 
renormalise according to \cite{Kaiser:2000gs}
$$
\t_R = Z^2 \t_B ~~~~~~~~~~~~~~~~~~~
1 + \L_{1R} = Z^2(1 + \L_{1B})
$$
where $Z$ is the usual multiplicative renormalisation factor for the 
axial current, $J_{\m5R}^0 = Z J_{\m5B}^0$.  In the KL chiral Lagrangian
formalism, the singlet field $\varphi^0$ is itself renormalised. Allowing
for a non-zero vacuum angle $\theta$, this field renormalisation is
$$
\varphi_R^0 = Z^{-1}\varphi^0 + \sqrt{2\over3}(Z^{-1}-1)\theta
$$} This means that while $\t$ coincides with $A$ (the non-perturbative parameter 
determining the QCD topological susceptibility) at $O(1)$, it differs beyond 
leading order in $1/N_c$.

The $SU(3)$ breaking which distinguishes the decay constants arises first 
from the terms in ${\cal L}_1$. Beyond this order, in addition to the direct 
contributions from the new couplings in the chiral Lagrangian, there are
also contributions from loop diagrams calculated using ${\cal L}_0$. These
give rise to the `chiral logarithms' $\m_P = {m_p^2\over32\pi^2 F^2}
\ln{m_P^2\over\m^2}$, which KL have calculated explicitly. 
To present their results, we again use the two-angle parametrisation 
(cf eq.{\ref{eq:dm}) in the octet-singlet sector:
\be
\left(\matrix{F^0_{\eta'} &F^0_{\eta}\cr F^8_{\eta'} &F^8_{\eta}}\right)
~~=~~\left(\matrix{F_0 \cos\vartheta_0 &-F_0 \sin\vartheta_0\cr
F_8 \sin\vartheta_8 &F_8 \cos\vartheta_8}\right)
\label{eq:ge}
\ee
To this order, the KL decay constants are then \cite{Kaiser:1998ds}:

\be
F_\pi ~=~ F\Bigl[1 + {4B\over F^2}\Bigl(2\sum m_q L_4 + 2m L_5\Bigr) ~+~
O(\m_P) \Bigr] 
\label{eq:gf}
\ee

\be 
F_K ~=~ F\Bigl[1 + {4B\over F^2}\Bigl(2\sum m_q L_4 + (m+m_s) L_5\Bigr) ~+~
O(\m_P) \Bigr] 
\label{eq:gg}  
\ee

\be
F_8 ~=~ F\Bigl[1 + {4B\over F^2}\Bigl(2\sum m_q L_4 + {2\over3}(m+2m_s) L_5
\Bigr) ~+~ O(\m_P)  \Bigr] 
\label{eq:gh}  
\ee

\noindent For the singlet,
\be
F_0 ~=~ \sqrt{1+\L_1} \bar F_0
\label{eq:gi}
\ee
where the scale-invariant part is
\be
\bar F_0 ~=~ F\Bigl[1+ {4B\over F^2}\Bigl(2\sum m_q L_4 + {2\over3}
(2m+m_s)(-L_5 + L_A)\Bigr) \Bigr]
\label{eq:gj}
\ee
where $L_A = (2L_5 + 3L_{18})/\sqrt{1+\L_1} = 2L_5 + O(1)$. Notice there are 
no loop corrections to $F_0$. Finally, the difference in the angles 
$\vartheta_0$ and $\vartheta_8$ is determined from
\begin{eqnarray}
&F^0_{\eta'} F^8_{\eta'} + F^0_{\eta} F^8_{\eta} ~=~ - F_0 F_8 \sin(\vartheta_0
- \vartheta_8)~~~~~~~~~~~~~~~~~~~~~ \nonumber \\
&~~~~~~~~~~~~~~~~~~~~~~~=~ {8\sqrt2\over3}B (m-m_s) (2L_5 + 3L_{18})
\label{eq:gk}
\end{eqnarray}
and is proportional to the $SU(3)$ breaking $m,m_s$ mass difference.
Also recall \cite{Gasser:1983yg} that at lowest order, the pseudoscalar masses
are $m_\pi^2 = 2mB$, ~$m_K^2 = (m+m_s)B$, ~$m_{\eta}^2 = {2\over3}(m+2m_s)B$
~and $m_{\eta'}^2 = {2\over3}(2m+m_s)B$.

These decay constants satisfy a set of relations which are closely analogous, 
but not identical, to the DGMOR relations (\ref{eq:bn}),(\ref{eq:bo}) 
and (\ref{eq:bp}). In fact, up to terms involving chiral logarithms, 
the decay constants shown above satisfy (see also ref.\cite{Feldmann:1999uf})
\begin{eqnarray}
&\bar F^0_{\eta'} \bar F^0_{\eta'} ~+~ \bar F^0_{\eta} \bar F^0_{\eta} ~~=~~
(\bar F^0)^2 ~~=~~ {1\over3}(F_\pi^2 + 2 F_K^2) 
\label{eq:gl} \\
\nonumber\\
&\bar F^0_{\eta'} F^8_{\eta'} ~+~ \bar F^0_{\eta} F^8_{\eta} ~~=~~
-\bar F^0 F^8 \sin(\vartheta_0 - \vartheta_8) ~~=~~
{2\sqrt2\over3}(F_\pi^2 - F_K^2) 
\label{eq:gm} \\
\nonumber\\
&F^8_{\eta'} F^8_{\eta'} ~+~ F^8_{\eta} F^8_{\eta} ~~=~~ (F^8)^2 ~~=~~
-{1\over3}(F_\pi^2 - 4 F_K^2)
\label{eq:gn} 
\end{eqnarray}

The corresponding DGMOR relations, i.e.~including the appropriate pseudoscalar
mass terms, are broken also by the terms proportional to $L_7$ and $L_8$
(the $L_6$ contributions cancel).
These are just the expected $O(m^2)$ corrections to the leading-order
PCAC relations.

Our main interest, as always, is in the flavour singlet sector. Notice that
the relation (\ref{eq:gl}) is written for the scale-invariant part of the
singlet decay constants only, omitting the factor involving the OZI-violating
coupling $\L_1$. This does not involve the topological susceptibility in any
way and is not related to the Witten-Veneziano formula. This enters the 
formalism as follows. In the chiral limit and working to $O(\d)$, 
the relevant part of the chiral Lagrangian is just
\be
{\cal L} ~\sim~ {1\over8}F^2(1+\L_1) \pl_\m\varphi^0 \pl^\m \varphi^0 ~-~ 
{3\over4} \t (\varphi^0)^2
\label{eq:go}
\ee 
from which it follows immediately that
\be
(F^0_{\eta'})^2 m_{\eta'}^2  ~=~ 6\t, ~~~~~~~~~~~
F^0_{\eta'} = \sqrt{1+\L_1}F
\label{eq:gp}
\ee
At leading order in $1/N_c$, this is the original Witten-Veneziano
formula, since $\L_1 = O(1/N_c)$ and we may interpret 
$\t = -\chi(0)|_{YM} + O(1/N_c)$.
However, beyond leading order and away from the chiral limit, it
does not have a straightforward generalisation in terms of the QCD
topological susceptibility, since $\t$ is not identical to $A$.
Fundamentally, this originates from the use of an RG non-invariant field 
$\varphi^0$ in the formulation of the chiral Lagrangian. In contrast, 
the corresponding formula (\ref{eq:bn}), i.e.~the flavour singlet DGMOR 
relation written with our RG invariant definition of $f^{0\eta'}$ and 
involving the parameter $A$ determining the topological susceptibility in 
QCD itself, is the appropriate generalisation of the Witten-Veneziano 
relation and provides a more suitable basis for describing $U(1)_A$ 
phenomenology.

These formulae allow a numerical determination of the decay constants 
and mixing angles in the octet-singlet sector. The results quoted in 
ref.\cite{Feldmann:1998vh} are
\begin{eqnarray}
&F_0 ~=~ 1.25 f_\pi ~~~~~~~~
F_8 ~=~ 1.28 f_\pi \nonumber\\
&~~~~\theta_0 ~=~ - 4 ~{\rm deg} ~~~~~~~~~
\theta_8  ~=~ - 20.5 ~{\rm deg}
\label{eq:gq}
\end{eqnarray}
These should be compared with our results, eq.(\ref{eq:dn}). Since the
definitions of the octet decay constants are the same, we expect 
$F_8 \simeq f_8$ and $\vartheta_8 \simeq \theta_8$. Indeed, the angles 
agree while $F_8$ is a little higher than $f_8$, by around 10\%. 
This is explicable by the fact that the KL fit incorporates the 
next-to-leading order corrections in the chiral expansion but not the 
radiative decays, whereas we have used the leading-order DGMOR relation
together with the octet radiative decay formula.
The definitions of the flavour singlet decay constant and mixing angle are
different in the two approaches, so cannot be directly compared.

Radiative decays are described in the chiral Lagrangian approach by the
Wess-Zumino-Witten term, which encodes the anomalous Ward identities.
The relevant part, constructed by Kaiser and Leutwyler \cite{Kaiser:2000gs},
is\footnote{In ref.\cite{Kaiser:2000gs},
the constants $\L_3$ and $\L_4$ are called $K_1$ and $K_2$ resp.  The
$\L_3$ notation is used in refs.\cite{Leutwyler:1997yr} and 
\cite{Feldmann:1999uf}. The radiative decay formulae (\ref{eq:gs}),
(\ref{eq:gt}) are identical to those quoted in ref.\cite{Feldmann:1999uf},
eqs.(39) rewritten in our notation.}
\be
{\cal L}_{\rm WZW} ~~=~~ - {\a N_c\over4\pi}\Bigl[
{\rm tr}(\hat e^2 \varphi) ~+~ {1\over3}\L_3{\rm tr}(\hat e^2){\rm tr}\varphi
~+~ 2B\L_4 {\rm tr}(\hat e^2 M\varphi) \Bigr]~F_{\m\n} \tilde F^{\m\n}
\label{eq:gr}
\ee
where $\varphi = \varphi^a T^a$ and $\hat e = {\rm diag}({2\over3}, -{1\over3}, 
-{1\over3})$ ~is the quark charge matrix. $\L_3$ is $O(1/N_c)$ and scales
in the same way as $\L_1$, while $\L_4$ is $O(1)$ and RG invariant. 
It is then straightforward to derive the following formulae (omitting the
$O(m)$ corrections arising from the $\L_4$ term for simplicity)
\be
F^0_{\eta'} g_{\eta'\c\c} ~+~ F^0_{\eta} g_{\eta\c\c} ~~=~~ 
(1 + \L_3)~ a^0_{\rm em}{\a\over\pi} 
\label{eq:gs}
\ee

\be
F^8_{\eta'} g_{\eta'\c\c} ~+~ F^8_{\eta} g_{\eta\c\c} ~~=~~
a^8_{\rm em}{\a\over\pi} 
\label{eq:gt}
\ee

These are the analogues of our eqs.(\ref{eq:be}) and (\ref{eq:bf}).
The key point is that in each case, the flavour singlet formula has to
incorporate OZI breaking by the inclusion of a new, process-specific
parameter, $\L_3$ in the KL formalism and $g_{G\c\c}$ in our approach,
which must be determined from the experimental data. In each case, this
removes the predictivity of the formula unless, as we have discussed, 
we can bring theoretical arguments to bear to argue that the OZI-violating
terms are small. In the KL case, this would presumably require the 
RG-invariant ratio $(1+\L_3)/(1+\L_1)$ to be close to 1. 
However, once again, we consider that the flavour singlet decay formula 
presented here has the added virtue of explicitly showing the link 
with the topological suceptibility and giving a physical interpretation 
of the new OZI-violating parameter in terms of the gluonic component 
of the physical $\eta'$ meson.


\vskip1.5cm

\begin{thebibliography}{999}

\bibitem{Shore:1999tw}
  G.~M.~Shore,
  Nucl.\ Phys.\ B {\bf 569} (2000) 107
  [arXiv:hep-ph/9908217].

\bibitem{Shore:1991pn}
  G.~M.~Shore and G.~Veneziano,
  Nucl.\ Phys.\ B {\bf 381} (1992) 3.

\bibitem{Gell-Mann:1968rz}
  M.~Gell-Mann, R.~J.~Oakes and B.~Renner,
  Phys.\ Rev.\  {\bf 175} (1968) 2195.

\bibitem{Dashen:1969eg}
  R.~F.~Dashen,
  Phys.\ Rev.\  {\bf 183} (1969) 1245.

\bibitem{Witten:1979vv}
  E.~Witten,
  Nucl.\ Phys.\ B {\bf 156} (1979) 269.

\bibitem{Veneziano:1979ec}
  G.~Veneziano,
  Nucl.\ Phys.\ B {\bf 159} (1979) 213.

\bibitem{Narison:1992fd}
  S.~Narison, G.~M.~Shore and G.~Veneziano,
  Nucl.\ Phys.\ B {\bf 391} (1993) 69.

\bibitem{Shore:1992pm}
  G.~M.~Shore and G.~Veneziano,
  Mod.\ Phys.\ Lett.\ A {\bf 8} (1993) 373.

\bibitem{Shore:2004cb}
  G.~M.~Shore,
  Nucl.\ Phys.\ B {\bf 712} (2005) 411
  [arXiv:hep-ph/0412192].

\bibitem{Goldberger:1958vp}
  M.~L.~Goldberger and S.~B.~Treiman,
  Phys.\ Rev.\  {\bf 111} (1958) 354.

\bibitem{Veneziano:1989ei}
  G.~Veneziano,
  Mod.\ Phys.\ Lett.\ A {\bf 4} (1989) 1605.

\bibitem{Shore:1990zu}
  G.~M.~Shore and G.~Veneziano,
  Phys.\ Lett.\ B {\bf 244} (1990) 75.

\bibitem{Shore:1991dv}
  G.~M.~Shore and G.~Veneziano,
  Nucl.\ Phys.\ B {\bf 381} (1992) 23.

\bibitem{Leutwyler:1997yr}
  H.~Leutwyler,
  Nucl.\ Phys.\ Proc.\ Suppl.\  {\bf 64} (1998) 223
  [arXiv:hep-ph/9709408].

\bibitem{Kaiser:1998ds}
  R.~Kaiser and H.~Leutwyler,
  arXiv:hep-ph/9806336.

\bibitem{Kaiser:2000gs}
  R.~Kaiser and H.~Leutwyler,
  Eur.\ Phys.\ J.\ C {\bf 17} (2000) 623
  [arXiv:hep-ph/0007101].

\bibitem{Gasser:1983yg}
  J.~Gasser and H.~Leutwyler,
  Annals Phys.\  {\bf 158} (1984) 142;~
  Nucl.\ Phys.\ B {\bf 250} (1985) 465.

\bibitem{Herrera-Siklody:1996pm}
  P.~Herrera-Siklody, J.~I.~Latorre, P.~Pascual and J.~Taron,
  Nucl.\ Phys.\ B {\bf 497} (1997) 345;~~
  Phys.\ Lett.\ B {\bf 419} (1998) 326

\bibitem{Feldmann:1999uf}
  T.~Feldmann,
  Int.\ J.\ Mod.\ Phys.\ A {\bf 15} (2000) 159
  [arXiv:hep-ph/9907491].

\bibitem{Shore:2001cs}
  G.~M.~Shore,
  Phys.\ Scripta {\bf T99} (2002) 84
  [arXiv:hep-ph/0111165].

\bibitem{Feldmann:1997vc}
  T.~Feldmann and P.~Kroll,
  Eur.\ Phys.\ J.\ C {\bf 5} (1998) 327
  [arXiv:hep-ph/9711231].

\bibitem{Feldmann:1998vh}
  T.~Feldmann, P.~Kroll and B.~Stech,
  Phys.\ Rev.\ D {\bf 58} (1998) 114006
  [arXiv:hep-ph/9802409].

\bibitem{DiVecchia:1980ve}
  P.~Di Vecchia and G.~Veneziano,
  Nucl.\ Phys.\ B {\bf 171} (1980) 253.

\bibitem{Veneziano:1990mx}
  G.~Veneziano,
Invited Talk Given at Okubofest: From Symmetries to Strings: 
Forty Years of Rochester Conferences, Rochester, N.Y., 1990;
CERN-TH-5840-90.

\bibitem{DelDebbio:2004ns}
  L.~Del Debbio, L.~Giusti and C.~Pica,
  Phys.\ Rev.\ Lett.\  {\bf 94} (2005) 032003
  [arXiv:hep-th/0407052].

\bibitem{Narison:1998aq}
  S.~Narison, G.~M.~Shore and G.~Veneziano,
  Nucl.\ Phys.\ B {\bf 546} (1999) 235
  [arXiv:hep-ph/9812333].

\bibitem{PDG}
  Particle Data Group, Review of Particle Properties, 
  Phys.\ Lett.\ {\bf B592} (2004) 1.

\bibitem{L3}
  M.~Acciarri {\it et al.}, L3 Collaboration,
  Phys.\ Lett.\ {\bf B418} (1998) 399.

\bibitem{Crystal}
  D.A.~Williams {\it et al.}, Crystal Ball Collaboration,
  Phys.\ Rev.\ {\bf D38} (1988) 1365.

\bibitem{ASP}
  N.A.~Roe {\it et al.}, ASP Collaboration,
  Phys.\ Rev.\ {\bf D41} (1990) 17.

\bibitem{PDG1994}
  Particle Data Group, Review of Particle Properties,
  Phys.\ Rev.\ {\bf D50} (1994) 1451.

\bibitem{DiGiacomo:1990ij}
  A.~Di Giacomo,
  Nucl.\ Phys.\ Proc.\ Suppl.\  {\bf 23B} (1991) 191.

\bibitem{Narison:1990cz}
  S.~Narison,
  Phys.\ Lett.\ B {\bf 255} (1991) 101;
  Z.\ Phys.\ C {\bf 26} (1984) 209.

\bibitem{Kochelev:2001nv}
  N.~I.~Kochelev and V.~Vento,
  Phys.\ Lett.\ B {\bf 515} (2001) 375;~
  {\it ibid.} {\bf 541} (2002) 281.

\bibitem{Shore:1998dn}
  G.~M.~Shore,
  {\it in} `From the Planck Length to the Hubble Radius', Erice 1998, 79-105.~
  arXiv:hep-ph/9812355.

\bibitem{Bass:2004xa}
  S.~D.~Bass,
  ``The spin structure of the proton,''
  arXiv:hep-ph/0411005.

\bibitem{Mallot}
  G.~Mallot, S.~Platchkov and A.~Magnon, CERN-SPSC-2005-017; SPSC-M-733.

\bibitem{Ageev:2005gh}
  E.~S.~Ageev {\it et al.}  [COMPASS Collaboration],
  Phys.\ Lett.\ B {\bf 612} (2005) 154
  [arXiv:hep-ex/0501073].

\bibitem{Ellis:1973kp}
  J.~R.~Ellis and R.~L.~Jaffe,
  Phys.\ Rev.\ D {\bf 9} (1974) 1444
  [Erratum-ibid.\ D {\bf 10} (1974) 1669].

\bibitem{Shore:1999be}
  G.~M.~Shore and B.~E.~White,
  Nucl.\ Phys.\ B {\bf 581} (2000) 409
  [arXiv:hep-ph/9912341].

\bibitem{Shore:2000ca}
  G.~M.~Shore,
  Nucl.\ Phys.\ Proc.\ Suppl.\  {\bf 96} (2001) 171
  [arXiv:hep-ph/0007239].

\bibitem{Bugg:2004cm}
  D.~V.~Bugg,
  Eur.\ Phys.\ J.\ C {\bf 33} (2004) 505.

\bibitem{Moskal:2004cm}
  P.~Moskal,
  ``Hadronic interaction of eta and eta$'$ mesons with protons,''
  arXiv:hep-ph/0408162.

\bibitem{Bass:2001ix}
  S.~D.~Bass,
  Phys.\ Scripta {\bf T99} (2002) 96
  [arXiv:hep-ph/0111180].

\bibitem{Moskal:2004nw}
  P.~Moskal {\it et al.},
  Int.\ J.\ Mod.\ Phys.\ A {\bf 20} (2005) 1880
  [arXiv:hep-ex/0411052].

\bibitem{Moskal:1998pc}
  P.~Moskal {\it et al.},
  Phys.\ Rev.\ Lett.\  {\bf 80} (1998) 3202
  [arXiv:nucl-ex/9803002].

\bibitem{Nakayama:2005ts}
  K.~Nakayama and H.~Haberzettl,
  ``Analyzing eta$'$ photoproduction data on the proton at energies of 1.5-GeV -
  2.3-GeV,''
  arXiv:nucl-th/0507044.

\bibitem{Dugger:2005du}
  M.~Dugger  [CLAS Collaboration],
  ``S=0 pseudoscalar meson photoproduction from the proton,''
  arXiv:nucl-ex/0512005.

\end{thebibliography}
\end{document}